\documentclass[aps,prb,superscriptaddress,reprint]{revtex4-1}

\usepackage{amsmath}
\usepackage{amssymb}
\usepackage{dsfont}
\usepackage{bm}
\usepackage[hidelinks]{hyperref}
\usepackage{graphicx}
\usepackage{subfigure}
\usepackage{color}
\usepackage{tikz}
\usepackage{relsize}

\newcommand{\ket}[1]{\mbox{$| #1 \rangle$}}
\newcommand{\ingr}[2]{\begin{matrix}\includegraphics[height=#1 cm]{#2}\end{matrix}}

\begin{document}
\title{Classification and detection of symmetry fractionalization in chiral spin liquids}
\author{Lukasz Cincio}
\affiliation{Perimeter Institute for Theoretical Physics, Waterloo, Ontario,
  N2L 2Y5, Canada} 
\author{Yang Qi}
\affiliation{Institute for Advanced Study, Tsinghua University,
  Beijing 100084, China}
\affiliation{Perimeter Institute for Theoretical Physics, Waterloo, Ontario,
  N2L 2Y5, Canada}

\begin{abstract}
  In this work we study the crystal symmetry fractionalization in
  chiral spin liquids with the chiral-semion topological order. We
  show that if such a chiral spin liquid is realized in a
  two-dimensional lattice model with odd number of spin-$\frac12$ per
  unit cell and the state preserves spin rotation symmetry and
  translation symmetries, the semion excitation must carry both
  half-integer spin and fractional crystal quantum numbers. As a
  result, only a unique symmetry enriched topological phase can be
  realized in chiral spin liquids in a spin-$\frac12$ kagome lattice
  model. These fractional symmetry quantum numbers are confirmed
  numerically using ground state wave functions obtained with the
  density matrix renormalization group method.
\end{abstract}

\maketitle

\section{Introduction}
\label{sec:introduction}

Quantum spin liquids~\cite{AndersonQSL, BalentsSLReview} with
topological orders~\cite{KivelsonPRB1987, WenTO1990, Wen1991a,
  WenZ2SL1991} have generated considerable interests in the study of
strongly correlated quantum systems, because they represent a class of
quantum phases beyond the Landau's paradigm of symmetry breaking
phases~\cite{landau}, and they are closely related to the study of
high-$T_c$ superconductivity~\cite{KivelsonTO1987, LNWReview}. So far
many different types of quantum spin liquids, including gapped and
gapless spin liquids, have been studied. Among these different types,
the chiral spin liquid state~\cite{KalmeyerCSL1987, WenCSL} was the
earliest one to be proposed theoretically. The chiral spin liquid
state has been constructed in exactly solvable model Hamiltonians~\cite{Schroeter2007}, and
it was recently found in a variety of spin-$\frac12$ models with
short-range interaction on the kagome lattice by different numerical
and theoretical approaches, including the density matrix
renormalization group (DMRG) method~\cite{Bauer2014, GongCSL2014,
  HeCSL2014, HePT2015, GongGPD2015}, fermionic parton constructions
and variational Monte Carlo method~\cite{MeiCSLX, HuCSL2015} and
theoretical constructions~\cite{Thomale2009,SunCSL}.

One question remains to be answered is whether the chiral spin liquid
states obtained in different methods are the same, i.e. whether they
belong to the same phase. This question can be easily answered for
topologically trivial phases because these phases are classified by
the symmetry the ground states have, and can be detected by measuring
correlation functions that detects all possible ways of symmetry
breaking~\cite{landau}. However, classifying and detecting topological
orders are much harder because these orders are not reflected in any
correlation functions of local observables.

First, the chiral spin liquid states have an intrinsic topological
order, which is characterized by topological excitations carrying
fractional statistics, and chiral central charges of the gapless edge
states. In particular, the chiral spin liquids have the
``chiral-semion'' topological order, which is the same as the
intrinsic topological order of a $\nu=\frac12$ bosonic fractional
quantum hall (FQH) state~\cite{LaughlinFQHE, KalmeyerCSL1987}. (In
this work, the terminology of ``chiral spin liquid'' only refers to
spin liquids with the chiral-semion topological order. In some
literatures~\cite{Messio2013, Bieri2015}, chiral spin liquids also refer to
time-reversal-symmetry-breaking spin liquids with $\mathbb Z_2$
topological order, and these are different from the chiral spin
liquids discussed in this work.) In this topological order there are
two types of anyons: the trivial anyon (denoted by $\mathds1$) and the
semion $s$. The semion $s$ has a fractional statistics such that
exchanging two semions gives a statistical phase of $e^{i\pi/2}=i$,
and it satisfies the fusion rule that two semions fuse into a trivial
anyon $s\times s=\mathds1$. The intrinsic topological order can be
detected numerically: the number of anyons can be measured from the
topological ground state degeneracy on a torus~\cite{WenGSD1989,
  WenNiuGSD1990} and topological entanglement
entropy~\cite{PriskillKitaev2006, LevinWen2006}, while the anyon
statistics can be measured from the modular
matrices~\cite{YZhangSTMat2012, Cincio2013, MeiSTMat2015}. These
methods have been applied to identify the chiral-semion topological
order in the DMRG studies~\cite{Cincio2013, GongCSL2014, HeCSL2014,
  HePT2015, GongGPD2015}.

Second, the crystal and spin rotation symmetries that are present in
the chiral spin liquid states further enriches their topological
order~\cite{XChenLUT}. Because there is only one type of nontrivial
anyon, the symmetry enriched topological (SET) order is classified by
the symmetry fractionalization carried by the semion
excitations~\cite{wenpsg, YaoFuQi2010X, Essin2013,
  MesarosSET2013}. Here, the symmetry fractionalization refers to the
phenomenon that an anyon excitation can carry a projective
representation of the symmetry group and consequently fractional
symmetry quantum numbers~\cite{Essin2013, MesarosSET2013, BarkeshliX,
  Fidkowski_unpub, Tarantino_arxiv, Teo2015, Lu2013,
  HungPRB2013}. This occurs in SET phases because, although any
physical state must carry a linear representation of the symmetry
group, it also contains a collection of anyons that fuse into
$\mathds1$. Therefore, each anyon can carry a projective
representation as long as the representations carried by the
collection of anyons always fuse into a linear representation. For
example, in a chiral spin liquid, the semions must appear in pairs in
a physical state because of the fusion rule $s\times s=\mathds1$.
Therefore, although each physical state must carry a linear
representation of the SO(3) spin rotation symmetry group, the semion
can carry a projective representation, and therefore carries a
half-integer spin.

The symmetry fractionalization can be classified by the projective
representation of the symmetry group according to the anyon fusion
rule~\cite{Essin2013, MesarosSET2013, BarkeshliX}. However, not all
the symmetry fractionalization obtained this way can be realized in a
two-dimensional (2D) system, as many of them are anomalous and can
only be realized on a surface of a three-dimensional (3D)
symmetry-protected topological (SPT) phase~\cite{XChenSPT}. For
on-site unitary symmetries, these anomalous symmetry
fractionalizations can be systematically identified~\cite{BarkeshliX,
  Chen2014}. However, for antiunitary and crystal symmetries such
general methods have not yet been established and anomalous symmetry
fractionalization has only been studied
case-by-case~\cite{VishwanathPRX2013, CWangETMT2013, Qi_unpub,
  HermeleFFAT}.

Since the information of symmetry fractionalization can distinguish
different symmetric spin liquid phases, the detection of symmetry
fractionalization is essential to the study of spin liquid
states. Methods of numerically and experimentally detecting the
symmetry fractionalization have been developed in the context of
$\mathbb Z_2$ spin liquids~\cite{wenpsg, MeiKagomeX, Essin2014,
  LWangSET2015, QiCSF, ZLVPSG}, and they can be generalized to be
applied to the chiral spin liquids.

In this work we answer the aforementioned question by studying the
crystal symmetry fractionalization in 2D chiral spin liquids with the
chiral-semion topological order. Particularly we find there is only
one way to fractionalize the crystal symmetries, and all the chiral
spin liquids found in previous studies must belong to the same
phase. We first discuss in general how to define fractionalized
crystal symmetry quantum numbers for anyons with fractional statistics
in Sec.~\ref{sec:csf-anyon}. We then classify crystal symmetry
fractionalization in 2D chiral spin liquids in
Sec.~\ref{sec:classify-csf}. In this work, we consider
spin-rotational-invariant chiral spin liquids with odd number of
spin-$\frac12$ per unit cell (this includes all the chiral spin
liquids found in different lattice models on the kagome
lattice). Under this assumption, there is only one nonanomalous
combination of fractional symmetry quantum number the semion can
take. Therefore, there is only one type of symmetric chiral spin
liquid that can be realized in a spin-$\frac12$ model on the kagome
lattice. In Sec.~\ref{sec:detect-dmrg}, we numerically study the
symmetry fractionalization in ground state wave functions obtained
using the DMRG method, and confirm that the chiral spin liquid state
indeed has the symmetry fractionalization determined in
Sec.~\ref{sec:classify-csf}. We conclude in Sec.\ref{sec:conclusion}.

\section{Crystal symmetry fractionalization of anyons}
\label{sec:csf-anyon}

In this section we discuss the definition of crystal symmetry
fractionalization for anyons carrying fractional statistics, and pave
the way for studying the classification of chiral spin liquids with
crystal symmetries in the rest of the paper. Here, we restrict
ourselves to Abelian topological orders.

The concept of crystal symmetry fractionalization was first
studied~\cite{wenpsg} in the context of analyzing the projective
symmetry group (PSG) of mean field ansatz in parton constructions,
where the partons carry a projective representation of the crystal
symmetry group, and thus fractional crystal symmetry quantum
numbers. Although the PSG analysis is useful in classifying symmetric
spin liquids in parton constructions (including gapless spin liquids
that are beyond the scope of gapped SET phases), it has limitations
when being used to classify symmetry fractionalization in SET
phases. First, the parton used in a construction usually corresponds
to only one type of anyon, so it does not explicitly give the full
symmetry fractionalization if there are more than one type of anyons
(this is not an issue for the chiral-semion topological
order). Second, the projective symmetry representation carried by the
parton may not be the same as the projective representation of the
corresponding anyon, as we shall demonstrate with an explicit example
in Appendix~\ref{sec:psg-af}.

For SET phases with the $\mathbb Z_2$ (toric code) topological order,
previous works~\cite{Essin2013, LuBFU, QiCSF, ZLVPSG} have provided
practical definitions of crystal symmetry fractionalization and
methods to measure it from the ground state wave functions. Here we
briefly review these definitions and measurements, and generalize them
to anyons taking fractional statistics other than the Bose and Fermi
statistics, like the semions in the chiral-semion topological order.

\begin{figure}[!t]
  \centering
  \subfigure[\label{fig:anyons:au:flat}]{\includegraphics[scale=0.9]{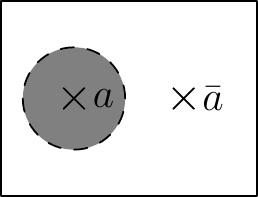}}~~~~~~
  \subfigure[\label{fig:anyons:au:sphere}]{\includegraphics[scale=0.9]{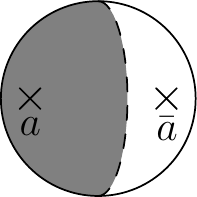}}~~~~~~
  \subfigure[\label{fig:anyons:au:tube}]{\includegraphics[scale=0.9]{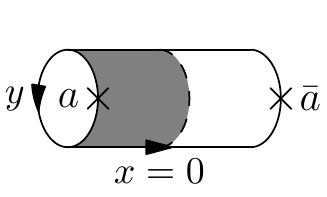}}\\
  \subfigure[\label{fig:anyons:u:flat}]{\includegraphics[scale=0.9]{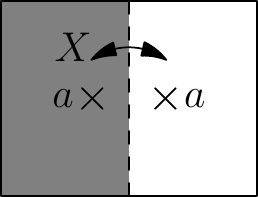}}~~~~~~
  \subfigure[\label{fig:anyons:u:sphere}]{\includegraphics[scale=0.9]{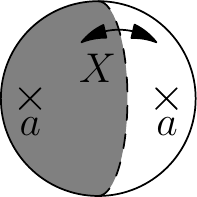}}~~~~~~
  \subfigure[\label{fig:anyons:u:tube}]{\includegraphics[scale=0.9]{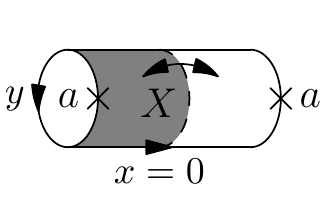}}
  \caption{(Color online) Symmetry actions on anyons. The ``x'' symbols mark the
    locations of anyons. The shaded area and the unshaded area mark
    the two regions in the Schmidt decomposition in
    Eqs.~\eqref{eq:Schmidt} and \eqref{eq:SchmidtLR}. In
    Figs. (a)-(c), an antiunitary symmetry $X$ maps $a$ and $\bar a$ to
    themselves; in Figs. (d)-(f), a unitary symmetry $X$ maps $a$ to
    $\bar a$ and vise versa. (a) A state that contains two anyons $a$ and
    $\bar a$ is shown. Each anyon is symmetric under an antiunitary
    symmetry $X$. (b) The configuration in (a) is smoothly deformed to
    a sphere. (c) The configuration in (b) is smoothly deformed to an
    infinite cylinder. (d) A configuration containing two anyons: $a$
    and $\bar a$, which are mapped into each other by a unitary
    symmetry $X$. (b) The configuration in (a) is smoothly deformed to
    a sphere. (c) The configuration in (b) is smoothly deformed to an
    infinite cylinder.}
  \label{fig:anyons}
\end{figure}

The key difficulty in defining the projective symmetry representation
carried by an anyon is that the anyon cannot appear alone in a
physical wave function. Instead, a physical wave function must contain
anyons which can be fused together as a trivial anyon. For example, in
the chiral-semion phase the semions must appear in pairs as
$s\times s=\mathds1$. Here we take the assumption of symmetry
localization~\cite{Essin2013, BarkeshliX}, which states that the
symmetry action on a physical wave function, which is always well
defined, can be fractionalized into actions localized in the
neighborhoods of the anyons. To define the symmetry action on an
individual anyon, we take a wave function $|\Psi\rangle$ containing
two well-separated anyons $a$ and $\bar a$ that fuses into the trivial
particle $a\times\bar a=\mathds1$. Then, we construct a Schmidt
decomposition between the region containing $a$ and the region
containing $\bar a$ [see Fig.~\ref{fig:anyons:au:flat}],
\begin{equation}
  \label{eq:Schmidt}
  |\Psi\rangle=\sum_\alpha\lambda_\alpha|a(\bm r)\rangle_\alpha
  \otimes|\bar a(\infty)\rangle_\alpha.
\end{equation}
In this Schmidt decomposition, $|a(\bm r)\rangle_\alpha$ are the
Schmidt eigenstates supported by the region containing an anyon $a$ at
$\bm r$, and $|\bar{a}(\infty)\rangle_\alpha$ are the Schmidt eigenstates
supported by the region containing an anyon $\bar a$ at infinity. The
symmetry action on the anyon can then be described by how the Schmidt
eigenstates transform under a symmetry operation $X$,
\begin{equation}
  \label{eq:grep}
  X: |a(\bm r)\rangle_\alpha\rightarrow U^X_{\alpha\beta}
  |a(\bm r)\rangle_\beta,
\end{equation}
where $U^X$ is the symmetry representation carried by the Schmidt
eigenstates $|a(\bm r)\rangle_\alpha$. It is a projective
representation with U(1) coefficients because of the U(1) gauge degree
of freedom present in the Schmidt decomposition in
Eq.~\eqref{eq:Schmidt}. Here we assume that a) a symmetry breaking
field is applied to the neighborhood of $a$ and $\bar a$ to lift any
symmetry-protected degeneracy of the anyons, and b) an infinitesimal
symmetry breaking field is applied near the boundary of the
decomposition to lift any symmetry-protected degeneracy in the
entanglement spectrum. Since $a$ is an Abelian anyon, the Schmidt
decomposition in Eq.~\eqref{eq:SchmidtLR} does not have any degeneracy
after all symmetries are broken. Therefore, the gauge degree of
freedom in the Schmidt decomposition is U(1).

This projective representation can be directly used to define the
quantum number fractionalization~\cite{QiCSF} $X^2=\pm1$ for an
antiunitary $\mathbb Z_2$ symmetry operation $X$ (like time-reversal). This is because the arbitrary U(1) phase factor in the
projective representation is canceled when $X$ is applied twice. In other words, for an
antiunitary group $\mathbb Z_2^X$,
$H^2(\mathbb Z_2^X, U(1))=H^2(\mathbb Z_2^X, \mathbb Z_2)$ . Here,
$H^2(G, R)$ denotes the second group cohomology with coefficient
$R$~\cite{XChenSPT}. The commutation relation
fractionalization~\cite{QiCSF} between two unitary symmetry operators
can also be defined in this way.

However, this projective representation cannot be used directly to
define the quantum number fractionalization $X^2=\pm1$ for a unitary
$\mathbb Z_2$ symmetry $X$ like inversion and mirror reflection,
because, due to the arbitrary U(1) phase factor, the value of $X^2$ can
smoothly interpolate between $+1$ and $-1$. In other words, $H^2(\mathbb Z_2^X, U(1))$ is always trivial for a unitary group.

Instead, this symmetry fractionalization can be defined using the
symmetry eigenvalue of a wave function containing two anyons located
at two symmetry-related positions~\cite{LuBFU, QiCSF}, as shown in
Fig.~\ref{fig:anyons:u:flat}. For anyons in the $\mathbb Z_2$
topological order, the symmetry fractionalization of a bosonic anyon
is defined simply using the symmetry parity eigenvalue of the
two-anyon wave function: the bosonic anyon carries trivial
(nontrivial) symmetry fractionalization if the two-anyon wave function
is symmetric (antisymmetric) under the symmetry,
respectively. However, for fermionic anyons an opposite definition was
adapted: the fermionic anyon carries trivial (nontrivial) symmetry
fractionalization if the two-anyon wave function is antisymmetric
(symmetric), because for a physical fermion without any symmetry
fractionalization the two-fermion wave function would be
antisymmetric. For an anyon with fractional statistics (like a semion)
there is no natural expectation of the parity of the two-anyon wave
function. For consistency, we choose to generalize the definition for
bosonic anyons to general Abelian anyons. In other words, regardless
of the self-statistics of the anyon, we always define the symmetry
fractionalization to be trivial (nontrivial) if the two-anyon wave
function is symmetric (antisymmetric) under the symmetry $X$.

The symmetry fractionalization defined above for both antiunitary and
unitary $\mathbb Z_2$ symmetries can be directly related to 1D SPT
invariants if the system is put on a infinite cylinder and viewed as a
quasi-1D system. This is discussed in the context of the $\mathbb Z_2$
topological order by~\citet{ZLVPSG}, and it can be generalized to
other topological orders. This relation to 1D SPT invariants will be
used in Sec.~\ref{sec:classify-csf} to identify anomalous combinations
of fractional symmetry quantum numbers, and in
Sec.~\ref{sec:detect-dmrg} to numerically detect the symmetry
fractionalization from ground state wave functions.

On an infinite cylinder, the ground states of a topologically ordered
system with $n$ types of anyons are $n$-fold degenerate (the
degeneracy is only approximate when the width of the cylinder is
finite). In the subspace formed by the degenerate ground states, a
special basis called the minimum entropy states~\cite{YZhangSTMat2012}
can be chosen. In this basis, each ground state corresponds to one
type of anyon flux threading through the cylinder. It is
denoted by $|a\rangle$, where $a$ labels the type of anyon. Viewed as
a quasi-1D system, the state $|a\rangle$ is a trivial (nontrivial) SPT
state protected by a $\mathbb Z_2$ symmetry $X$ if the anyon $a$
carries a trivial (nontrivial) symmetry fractionalization $X^2=\pm1$,
respectively. The way the crystal symmetry $X$ is set up depends on
whether $X$ is unitary or antiunitary: an antiunitary symmetry should
be implemented such that it preserves the longitudinal direction
$Xx=x$, while a unitary symmetry should reverse the longitudinal
direction $Xx=-x$, where $x$ denotes the longitudinal coordinate, as
shown in Fig.~\ref{fig:anyons}. These setups are consistent with the
fact that both on-site antiunitary symmetries and unitary reflection
symmetries protect nontrivial 1D SPT states, and the SPT invariant is
encoded in the way the symmetry acts on the Schmidt eigenstates
obtained from a Schmidt decomposition of the ground state, when the
cylinder is cut into left and right parts at
$x=0$~\cite{Pollmann2010},
\begin{equation}
  \label{eq:SchmidtLR}
  |\Psi\rangle=\sum_\alpha\lambda_\alpha|L\rangle_\alpha
  \otimes|R\rangle_\alpha.
\end{equation}

For an antiunitary symmetry, the Schmidt decomposition in
Eq.~\eqref{eq:SchmidtLR} can be smoothly deformed to the one in
Eq.~\eqref{eq:Schmidt}, as shown in Fig.~\ref{fig:anyons:au:tube}, and
both the 1D SPT and symmetry fractionalization is defined using the
double action of $X$ on the left Schmidt eigenstates.

For a unitary symmetry, the symmetry operator $X$ maps the left part
to the right part and vise versa. This mapping is symmetric
(antisymmetric) if the state is a trivial (nontrivial) SPT. If we
consider instead a finite cylinder with two open boundaries on both
sides [see Fig.~\ref{fig:anyons:u:tube}], each boundary then hosts an
anyon $a$, and it can be shown that the $X$-parity of this two-anyon
wave function is determined by how $X$ maps between left and right
Schmidt states. If the mapping is symmetric (antisymmetric), the wave
function is even (odd) under $X$. Therefore, according to our
definition, the anyon carries a trivial (nontrivial) symmetry
fractionalization of $X^2=\pm1$ if $|a\rangle$ is a trivial
(nontrivial) 1D SPT, respectively.

We note that our definition of symmetry fractionalization for a
unitary $\mathbb Z_2$ symmetry is consistent regardless of the anyon
statistics, and it is opposite to the definition previously used for
fermionic anyons~\cite{Essin2013, LuBFU, QiCSF}. It is also different
from the projective representation of a fermionic parton
operator~\cite{wenpsg, LuBFU, QiCSF}. Using this definition, the
classification of symmetry fractionalization always directly maps to
1D SPT classification, where a trivial (nontrivial) symmetry
fractionalization corresponds to a trivial (nontrivial) 1D
SPT. Furthermore using this definition it is easy to derive the
fractional quantum number of an anyon bound state using the fusion
rule. If two anyons fuse into a third anyon $a\times b=c$, the
fractional quantum number of $c$, $X^2_c$, is directly obtained by
multiplying the fractional quantum numbers of $a$ and $b$, $X^2_a$ and
$X^2_b$:
\begin{equation}
  \label{eq:abc}
  X^2_c=X^2_a X^2_b,
\end{equation}
without the need to include a twist factor depending on the anyon
statistics~\cite{Essin2013, LuBFU, QiCSF}. Although we will not use
Eq.~\eqref{eq:abc} in this work (because there is only one type of
nontrivial anyon in the chiral-semion topological order), this fusion
rule and our definition of anyon symmetry fractionalization will be
useful in the study of crystal symmetry fractionalization in more
complicated topological orders, like the double-semion topological
order.

\section{Classification of 2D chiral spin liquid}
\label{sec:classify-csf}

Using the definition of crystal symmetry fractionalization described
in the previous section, here we study the classification of crystal
symmetry fractionalization in a 2D chiral spin liquid with the
chiral-semion topological order. The key result of this section is the following: if the semion excitation carries a half-integer spin, it must
also carry fractionalized crystal symmetry quantum
numbers. In particular, only one type of fully symmetric chiral spin
liquid can be realized on a 2D kagome lattice.

We begin with enumerating the possible ways to fractionalize crystal
symmetries in a chiral spin liquid on the kagome lattice. In a
chiral spin liquid state the time-reversal and mirror symmetries are
broken by the spin chirality order parameter
$\bm S_i\cdot\left(\bm S_j\times\bm S_k\right)$, which is odd under
both time-reversal and mirror symmetries. However the combination of
mirror and time-reversal is still a symmetry operation. Therefore the
symmetry group is reduced from $G=p6m\times\mathbb Z_2^T$ (which is
generated by two translations $T_{1,2}$, two mirror reflections $\mu$
and $\sigma$, and time-reversal operation $T$, as shown in Figs.~\ref{fig:XC8-4} and \ref{fig:XC8}) to
$p6m^\ast$, which is the group with the identical structure as $p6m$
but the mirror operations $\mu$ and $\sigma$ are replaced by
antiunitary operators $\mu^\ast = \mu T$ and $\sigma^\ast=\sigma T$,
respectively.

\begin{table}[htbp]
  \centering
  \begin{tabular*}{\columnwidth}{@{\extracolsep{\fill}}ccc}
    \hline\hline
    Quantum number & Algebraic relation & Anomaly-free choice\\
    \hline
    $\omega_{12}$ & $T_1T_2=\pm T_2T_1$ & $-1$ \\
    $\omega_{\mu^\ast}$ & $(\mu^\ast)^2=\pm1$ & $-1$ \\
    $\omega_{\sigma^\ast}$ & $(\sigma^\ast)^2=\pm1$ & $-1$ \\
    $\omega_I$ & $I^2=\pm1$ & $-1$\\
    \hline\hline
  \end{tabular*}
  \caption{Quantum numbers labeling different projective
    representations of the symmetry group $p6m^\ast$.}
  \label{tab:omega}
\end{table}

Since the semions obey the fusion rule $s\times s=\mathds1$, the
possible projective representations of $p6m^\ast$ it can carry is
classified by
$H^2(p6m^\ast, \mathbb Z_2)=H^2(p6m, \mathbb Z_2)=\mathbb Z_2^4$, and
are therefore labeled by four $\mathbb Z_2$ variables~\cite{ZhengPSG}
$\omega_{12}$, $\omega_{\mu^\ast}$, $\omega_{\sigma^\ast}$ and
$\omega_I$, where $\omega_{12}$ denotes the commutation relation
fractionalization~\cite{QiCSF} $T_1T_2=\pm T_2T_1$ and the other three
variables $\omega_X$ denote quantum number fractionalization
$X^2=\pm1$, as listed in Table~\ref{tab:omega}.

These $2^4=16$ different ways of symmetry fractionalization listed in
Table~\ref{tab:omega} are all consistent with the semion fusion rule.
However, as we will show in the rest of the section, most of them are
anomalous and can only be realized on the 2D surface of a 3D SPT
state. In other words, for an SET order to be realizable in 2D models,
it needs to satisfy anomaly-free conditions which greatly restricts
possible ways to fractionalize crystal symmetries. First, if the
semion carries a half-integer spin representation, it must also carry
fractionalized crystal symmetry quantum numbers
$\omega_{\mu^\ast}=\omega_{\sigma^\ast}=\omega_I=-1$. Second, if the model
has odd number of spin-$\frac12$ per unit cell, the semion must carry
a half-integer spin and $\omega_{12}=-1$. Combining these two
arguments we can show that there is only one anomaly-free SET order
for a fully symmetric chiral spin liquid state on the kagome
lattice.

We begin with an argument showing that in an anomaly-free SET order in
which the semion carries half-integer spin, it must also carry
fractional quantum numbers
$\omega_{\mu^\ast}=\omega_{\sigma^\ast}=\omega_I=-1$. Here our
argument is based on the flux-fusion anomaly test developed
by~\citet{HermeleFFAT}, which can also be used to restrict crystal
symmetry fractionalization visons can carry in a $\mathbb Z_2$ spin
liquid~\cite{QiSFV2015X}. In this test, we put the chiral spin liquid
state on a quasi-1D torus as shown in
Figs.~\ref{fig:anyons:au:tube} and \ref{fig:anyons:u:tube}, where the
spin liquid state has two nearly degenerate ground states, each
corresponds to one type of anyon flux going through the torus. We
denote the two states with a trivial anyon flux and a semion flux as
$|\mathds 1\rangle$ and $|s\rangle$, respectively.

If the semion carries a half-integer spin, it implies that it carries
a half-integer charge of the U(1) subgroup generated by $S^z$. Hence
the ground state $|\mathds1\rangle$ can be smoothly turned to $|s\rangle$ by
adiabatically threading in a $2\pi$ flux of this U(1) subgroup along
the longitudinal direction in the Hamiltonian. If the Hamiltonian is a
Heisenberg model, threading a $\phi$ flux is achieved by changing the
off diagonal coupling to
$J_{ij} ( e^{i\phi\frac{y_i-y_j}{L_y}} S_i^+S_j^- + e^{-i\phi\frac{y_i-y_j}{L_y}} S_i^-S_j^+)$, where
$y_i$ is the transverse coordinate and $L_y$ is the circumference of the
torus in the transverse direction. For a general Hamiltonian, this is
achieved by applying the unitary transformation
\begin{equation}
  \label{eq:U}
  U(\phi) = \prod_j e^{i\phi \frac {y_j}{L_y}S_j^z}.  
\end{equation}
If $\phi$ is smoothly varied from 0 to $2\pi$, the two ground states
are adiabatically connected to each other. This method has
been applied in DMRG studies to compute the second ground state from the previously obtained one~\cite{He2014,Zhu2015,GongCSL2014,HeCSL2014}.

Now we consider a $\mathbb Z_2$ symmetry operation $X$ satisfying
$X^2=1$. As explained in Sec.~\ref{sec:csf-anyon}, if the semion $s$
carries a nontrivial symmetry fractionalization $X^2=-1$, the ground
state $|s\rangle$ belongs to a different SPT state protected by $X$,
as compared to $|1\rangle$. [This requires that $Xx= + x$ ($Xx = - x$) if $X$ is
antiunitary (unitary), respectively.] The key argument of this anomaly
test is that if a symmetry $X$ is preserved throughout the process of
flux insertion, the two ground states must belong to the same 1D SPT
phase protected by $X$. This in turn implies that the semion
carries a trivial symmetry fractionalization of $X$.

We begin with the example of $M^\ast = MT$, where $M$ is any mirror
operation (for example $M$ can be either $\mu$ or $\sigma$ on the
kagome lattice). Because $M^\ast$ is an antiunitary operation, we
need to set up the torus such that the mirror axis is along the
longitudinal direction. Since $T$ commutes with $U(\phi)$ and
$M$ maps $y$ to $-y$, $M^\ast$ maps $U(\phi)$ in Eq.~\eqref{eq:U} to
$M^\ast U(\phi)M^\ast = U(-\phi)$ and hence reverses the threaded
flux. Therefore the symmetry $M^\ast$ is not preserved in the process
of threading the flux.

However, the combined symmetry operation $M^\ast e^{i\pi S^x}$ (here
the operation $e^{i\pi S^x}$ denotes the global symmetry operation
$\prod_j e^{i\pi S_j^x}$) is preserved during the flux threading,
because $S^x$ anticommutes with $S^z$, and a $\pi$-rotation with
respective to $S^x$ again reverses the threaded flux. Consequently
from the anomaly test argument we conclude that the semion must carry
a trivial symmetry fractionalization
$\left(M^\ast e^{i\pi S^x}\right)^2=+1$.
Since the semion also carries a half-integer spin, it fulfills $\left(e^{i\pi S^x}\right)^2=-1$. Furthermore, the spin rotation $e^{i\pi S^x}$ must commute with $M^\ast$~\footnote{Here
  the action of $M^\ast$ and $e^{i\pi S^x}$ must commute in the
  projective representation carried by the semion, because in the spin
  rotation symmetry group SO(3), $e^{i\pi S^x}$ can be smoothly
  deformed to the identity element, which commutes with $M^\ast$.}. From this, we conclude that the semion must also carry a fractionalized quantum number of $(M^\ast)^2=-1$.

Next, we consider the inversion symmetry $I$. Similarly, $I$
commutes with $S^z$ and maps $y$ to $-y$. This means that inversion $I$ is not preserved during the flux threading. However, the combination $I e^{i\pi S^x}$ is preserved. Hence, the
semion must carry $\left(I e^{i\pi S^x}\right)^2=+1$ and consequently
$I^2=-1$. 

In summary, if the semion carries half-integer spin, it must also
carry fractionalized crystal symmetry quantum numbers $(M^\ast)^2=-1$
and $I^2=-1$. For example, on the kagome lattice, the semion must
carry $\omega_{\mu^\ast}=\omega_{\sigma^\ast}=\omega_I=-1$.

Furthermore, if the lattice model has odd number of spin-$\frac12$ per
unit cell, it can be shown that~\cite{Zaletel2015, ChengLSM} the
semion must carry a half-integer spin and $\omega_{12}=-1$. First,
semion, as the only type of nontrivial anyon, must carry a half-integer
spin to screen the background spin-$\frac12$ quantum number per unit
cell. Second, the screening requires a background anyon charge of odd
number of semions per unit cell. This means that if a semion moves around an unit
cell, it sees a $\pi$ flux due to the self-statistics of the
semion. Hence the semion carries $\omega_{12}=-1$. Therefore in
lattice models like spin-$\frac12$ models on the kagome lattice
(with full spin rotation symmetry) the crystal symmetry
fractionalization carried by the semion is completely fixed as
summarized in the last column of Table~\ref{tab:omega}. Consequently, there is
one possible SET phase of a symmetric chiral spin liquid.

\section{Detecting crystal symmetry fractionalization in DMRG results}
\label{sec:detect-dmrg}
In this section we provide numerical confirmation for the theoretical
arguments given in Sec.~\ref{sec:classify-csf}. Furthermore, the
method presented here can be applied to systems realizing different
topological orders or when the assumptions made in
Sec.~\ref{sec:classify-csf} are not met (for example, in models that lack the SU(2) spin-rotational invariance, or on lattices with an even
number of sites per unit cell).

\begin{figure}[!t]
	\begin{center}
		\includegraphics[width=\columnwidth]{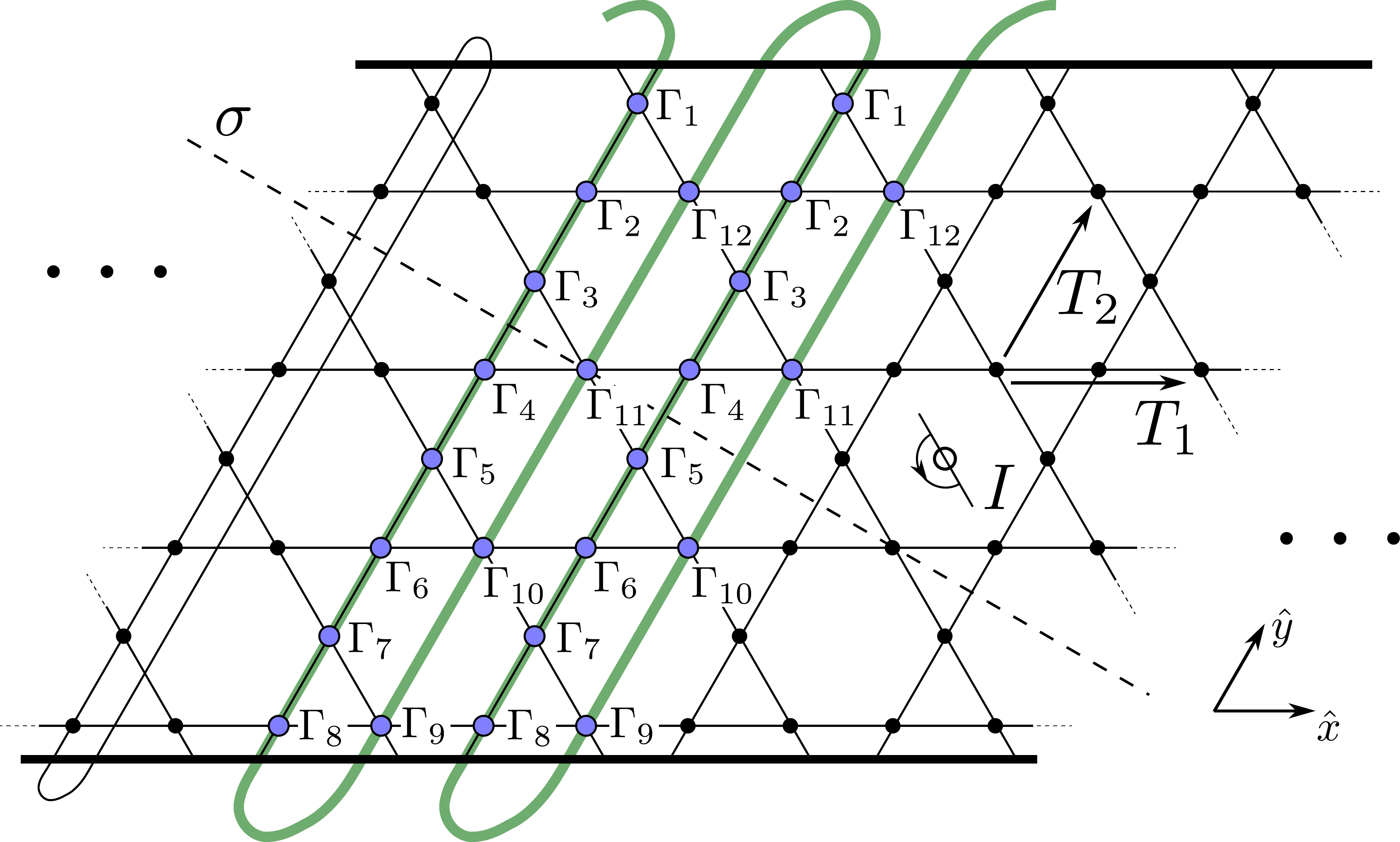}
	\end{center}
	\caption{(Color online) XC8-4 infinite cylinder used to extract $\omega_{12}$, $\omega_{\sigma^\ast}$ and $\omega_I$ quantum numbers. Semi-transparent green line represents an MPS path that covers the infinite cylinder. An MPS unit cell consists of $N=12$ tensors $\Gamma_j$ ordered along the path, see Eq.~(\ref{eq:MPSUnitCell}) for comparison. Matrices $\Lambda_j$ are omitted from the figure to improve clarity. An MPS unit cell (or its multiple) is compatible with translation $T_2$ and mirror reflection $\sigma$. Inversion $I$ maps an MPS unit cell to itself, provided it is ``shifted'' by two position along the path. All the symmetry operations act as permutations of lattice sites within (possibly enlarged) MPS unit cell.
	}
	\label{fig:XC8-4}
\end{figure}

\begin{figure}[!t]
	\begin{center}
		\includegraphics[width=\columnwidth]{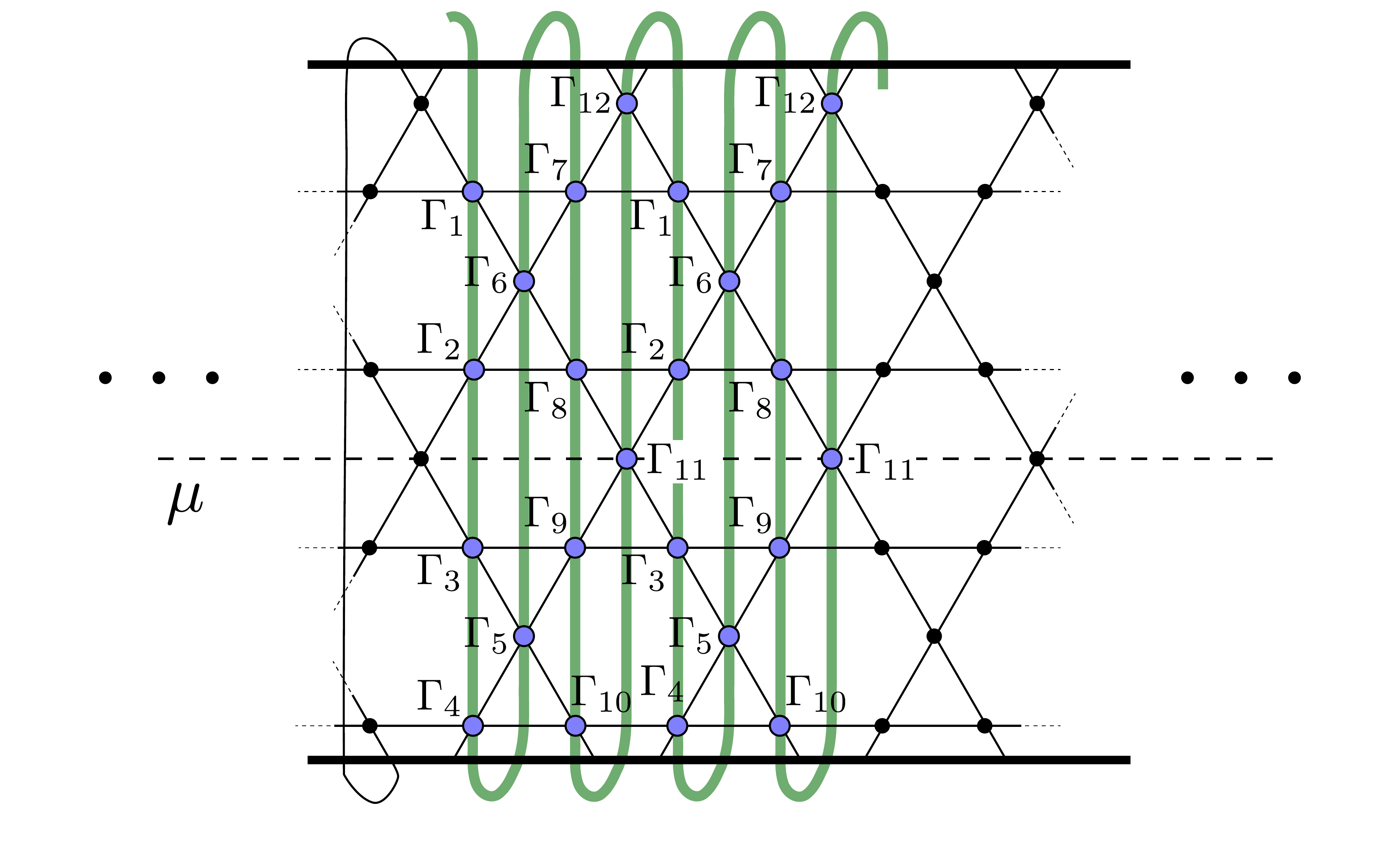}
	\end{center}
	\caption{(Color online) The computation of $\omega_{\mu^\ast}$ quantum number form an XC8 infinite cylinder. Mirror reflection $\mu$ maps an MPS unit cell to itself. An infinite cylinder is covered by an MPS path illustrated with a semi-transparent green line. Matrices $\Lambda_j$ are not shown fro simplicity.
	}
	\label{fig:XC8}
\end{figure}

For this purpose we study the following SU(2) invariant Hamiltonian on the kagome lattice
\begin{equation} \label{eq:Ham}
	H = \cos \theta \sum_{\langle i,j \rangle} \bm S_i \cdot \bm S_j \ + \ 
	\sin \theta \sum_{i,j,k \in \bigtriangleup, \bigtriangledown} \bm S_i \cdot (\bm S_j \times \bm S_k) \ ,
\end{equation}
where $\bm S_i$ is a spin-$\frac12$ operator acting on site $i$ of the
lattice, $\langle \ldots \rangle$ denotes nearest neighbors and sites
$i,j,k$ in the second sum are ordered clockwise around every
elementary triangle of the kagome lattice. The Hamiltonian (for
non-zero $\theta$) breaks time-reversal $T$ and mirror symmetries
$\sigma$, $\mu$ but remains invariant under combinations $\sigma^\ast
= \sigma T$, $\mu^\ast = \mu T$. It is also invariant under the
reflection $I$ with respect to the center of the hexagon (see
Figs.~\ref{fig:XC8-4} and \ref{fig:XC8}). Hence, the Hamiltonian has the $p6m^\ast$ symmetry discussed
in Sec.~\ref{sec:classify-csf}.

Ref.~\onlinecite{Bauer2014} provides compelling evidence that the
Hamiltonian in Eq.~(\ref{eq:Ham}), considered with
$0.05\pi \leq \theta \leq 0.5\pi$, realizes chiral $\mathbb{Z}_2$
semion topological order. In our present study we take one particular
value of the parameter $\theta=0.2\pi$ for concreteness. The
identification of an intrinsic topological order was performed by analyzing two
ground states on an infinite cylinder found by DMRG~\cite{Bauer2014}. Naturally, these
ground states found by DMRG are the minimum entropy
states~\cite{YZhangSTMat2012} $|\mathds1\rangle$ and
$|s\rangle$, which has an anyon flux of $\mathds1$
and $s$ threading through the cylinder, respectively. The anyon flux
threading through the cylinder in each ground state can be determined
by studying the SU(2) symmetry properties of the corresponding
entanglement spetrum~\cite{Cincio2013,Bauer2014}. Namely, the ground
state $\ket{\mathds1}$ ($\ket{s}$) gives rise to the entanglement
spectrum that transforms under integer (half-integer) representations
of SU(2). This characterization identifies $\ket{s}$ from the two
ground states, which will be used here to further measure the symmetry
fractionalization.

Our goal here is to extract all four quantum numbers $\omega_{12}$,
$\omega_{\sigma^\ast}$, $\omega_{\mu^\ast}$ and $\omega_I$ from the
ground state wave function $|s\rangle$ given by DMRG. It is done by
computing their corresponding 1D SPT
invariants\cite{Pollmann2012,ZLVPSG}, as detailed below. In order to
compute $\omega_{12}$, $\omega_{\sigma^\ast}$, $\omega_I$ we find
$| s \rangle$ on an XC8-4 infinite cylinder (see
Fig.~\ref{fig:XC8-4}), while for the calculation of $\omega_{\mu^\ast}$ we resort
to the ground state on XC8 infinite cylinder shown on Fig.~\ref{fig:XC8}. Here we use the
nomenclature introduced in Ref.~\onlinecite{Yan11}.

Let us consider the ground state $|s\rangle$ as a strictly 1D translationally invariant state on an infinite chain. Its
matrix-product representation reads
\begin{equation}
	\label{eq:mps}
	|s\rangle \ = \ \ingr{1.0}{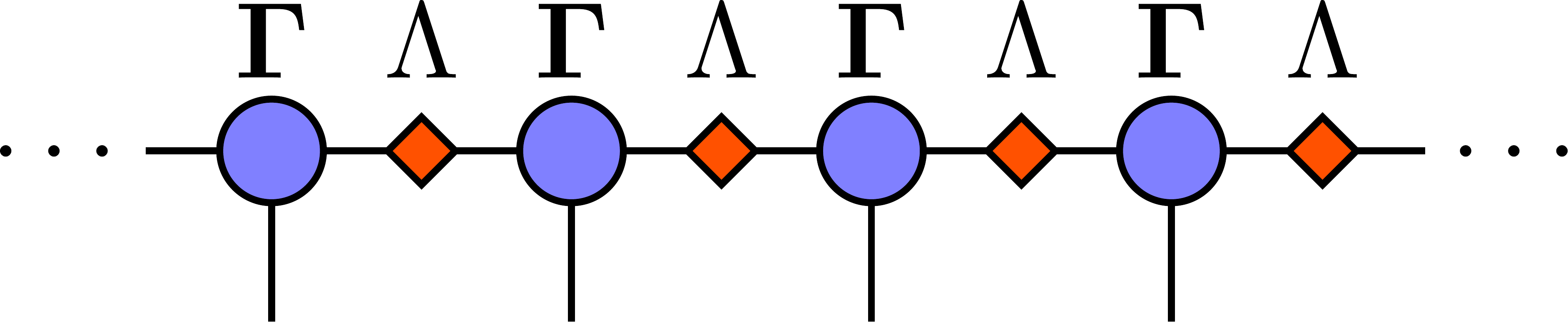} \ ,
\end{equation}
where rank-3 tensor $\bm \Gamma$ and matrix $\Lambda$ are chosen to fulfill canonical relations
\begin{equation}
	\label{eq:canonical}
	\ingr{1.694947}{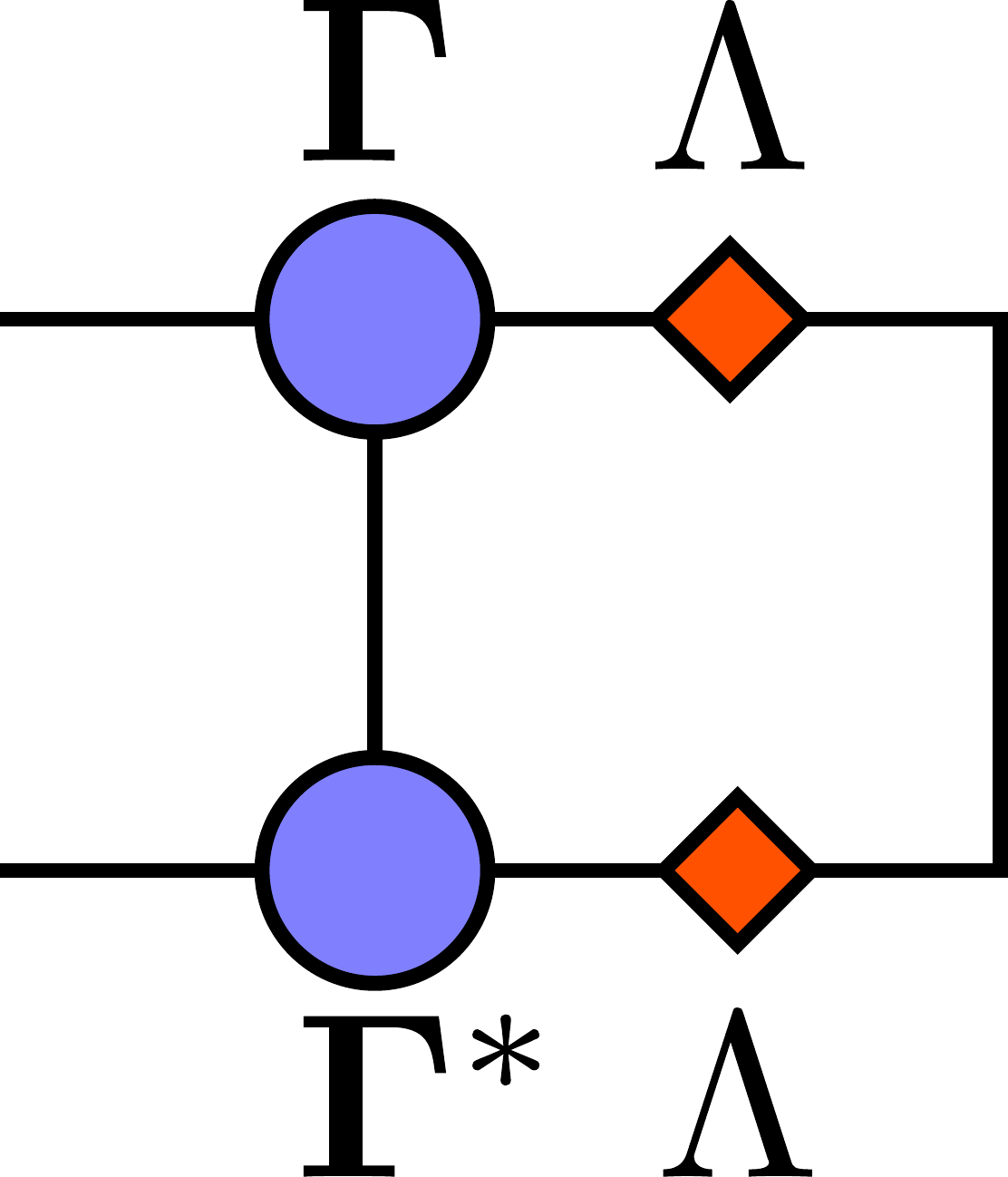} \ = \ \ingr{1.694947}{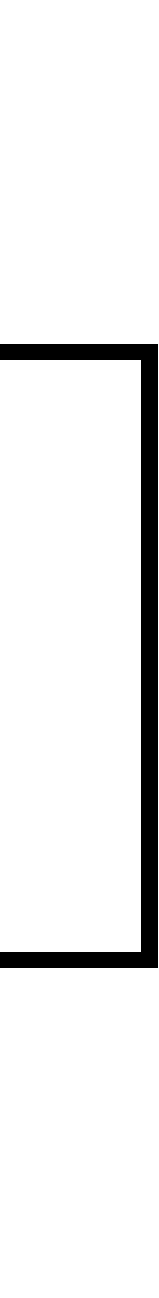} \ ,\quad 
	\ingr{1.694947}{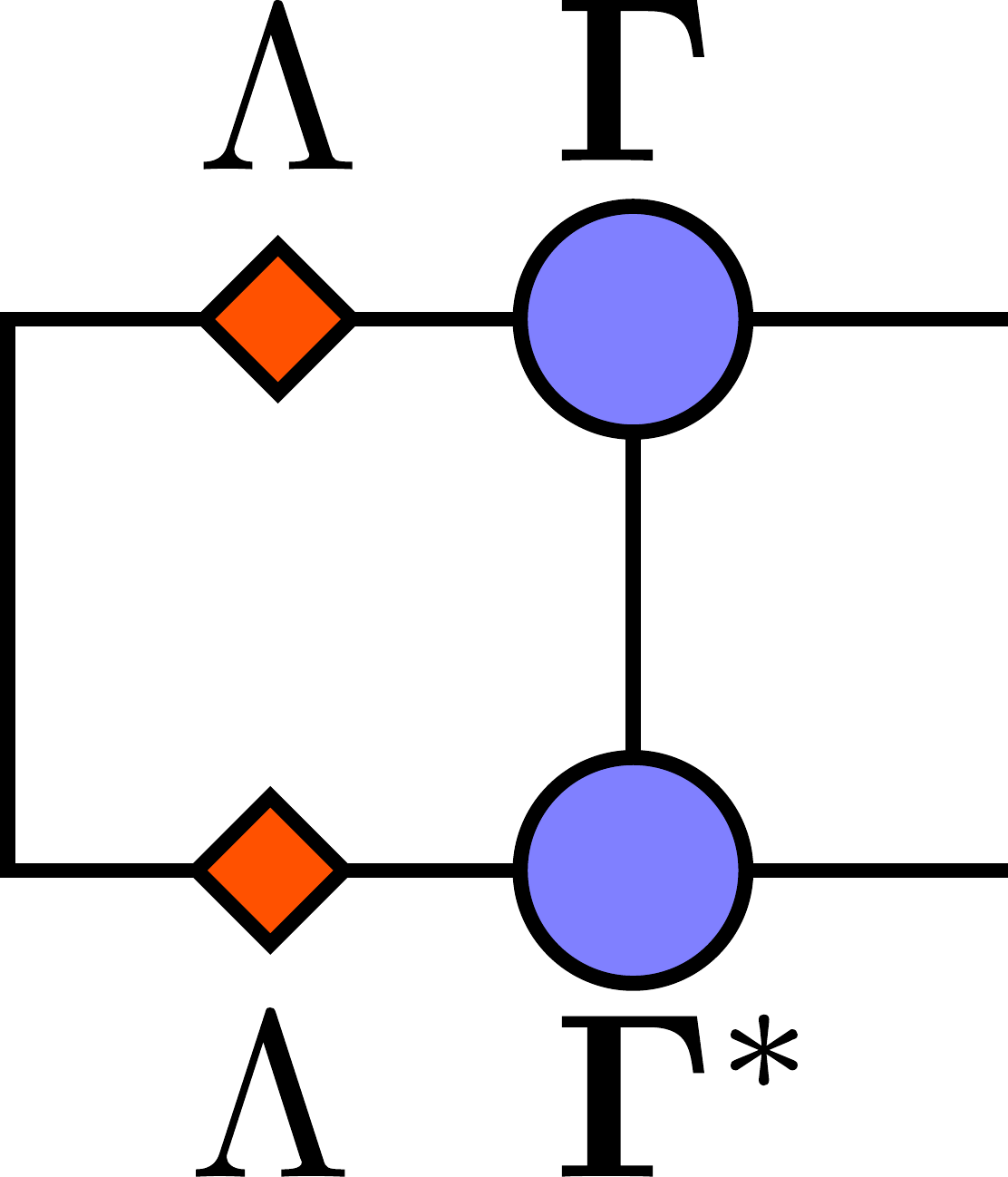} \ = \ \ingr{1.694947}{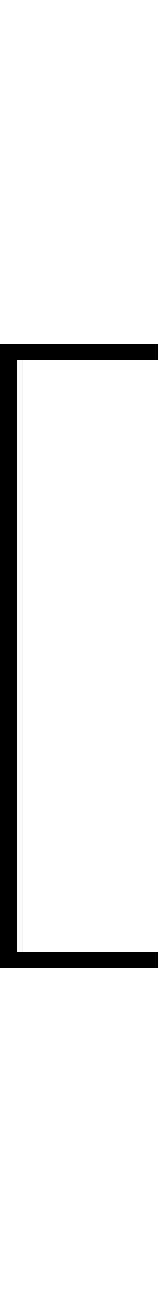} \ .
\end{equation}
Here, $\Lambda$ is a diagonal matrix with entries $\Lambda_{\alpha,\alpha} = \lambda_\alpha$, where $\lambda_\alpha$ are the Schmidt coefficients introduced in Eq.~(\ref{eq:SchmidtLR}). Tensor $\bm \Gamma$ represents a matrix-product state (MPS) unit cell that is a result of covering an infinite cylinder
\begin{equation}
	\label{eq:MPSUnitCell}
	\ingr{1.0}{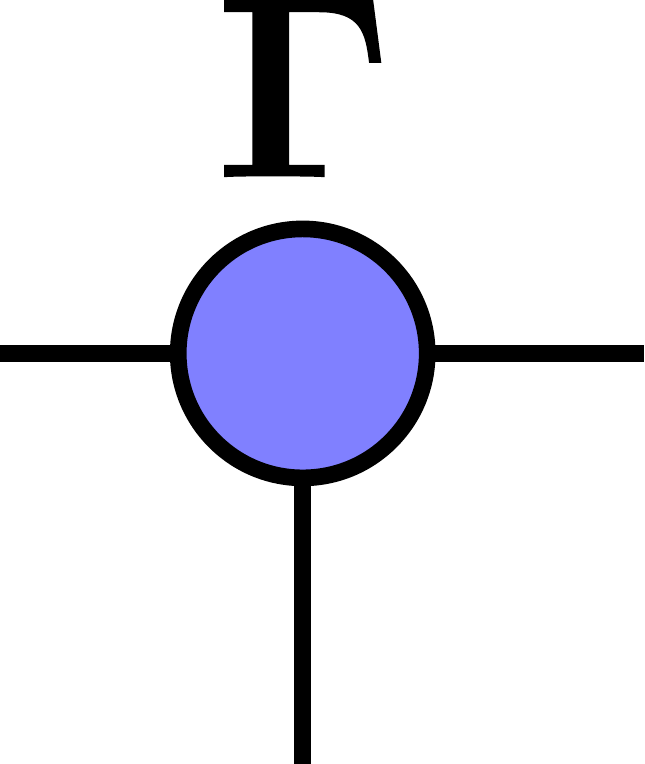} \ = \ \ingr{1.0}{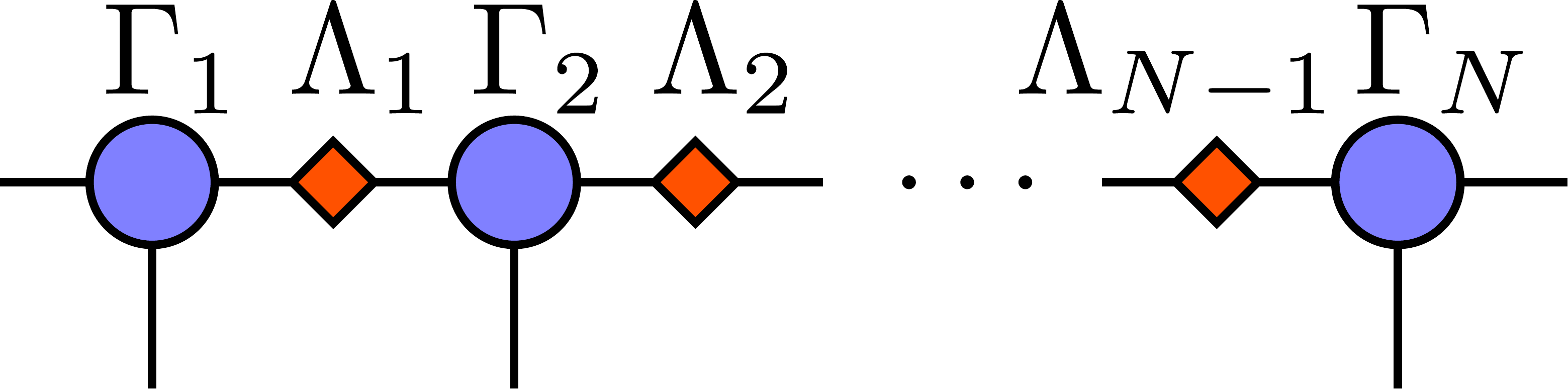} \ ,
\end{equation}
where $N$ is the size of an MPS unit cell.

It follows from Eq.~(\ref{eq:canonical}) that the identity $\bm 1$ is the right eigenvector of a transfer matrix
\begin{equation}
	\label{eq:TM}
	\mathcal{T} \ \equiv \ \ingr{1.694947}{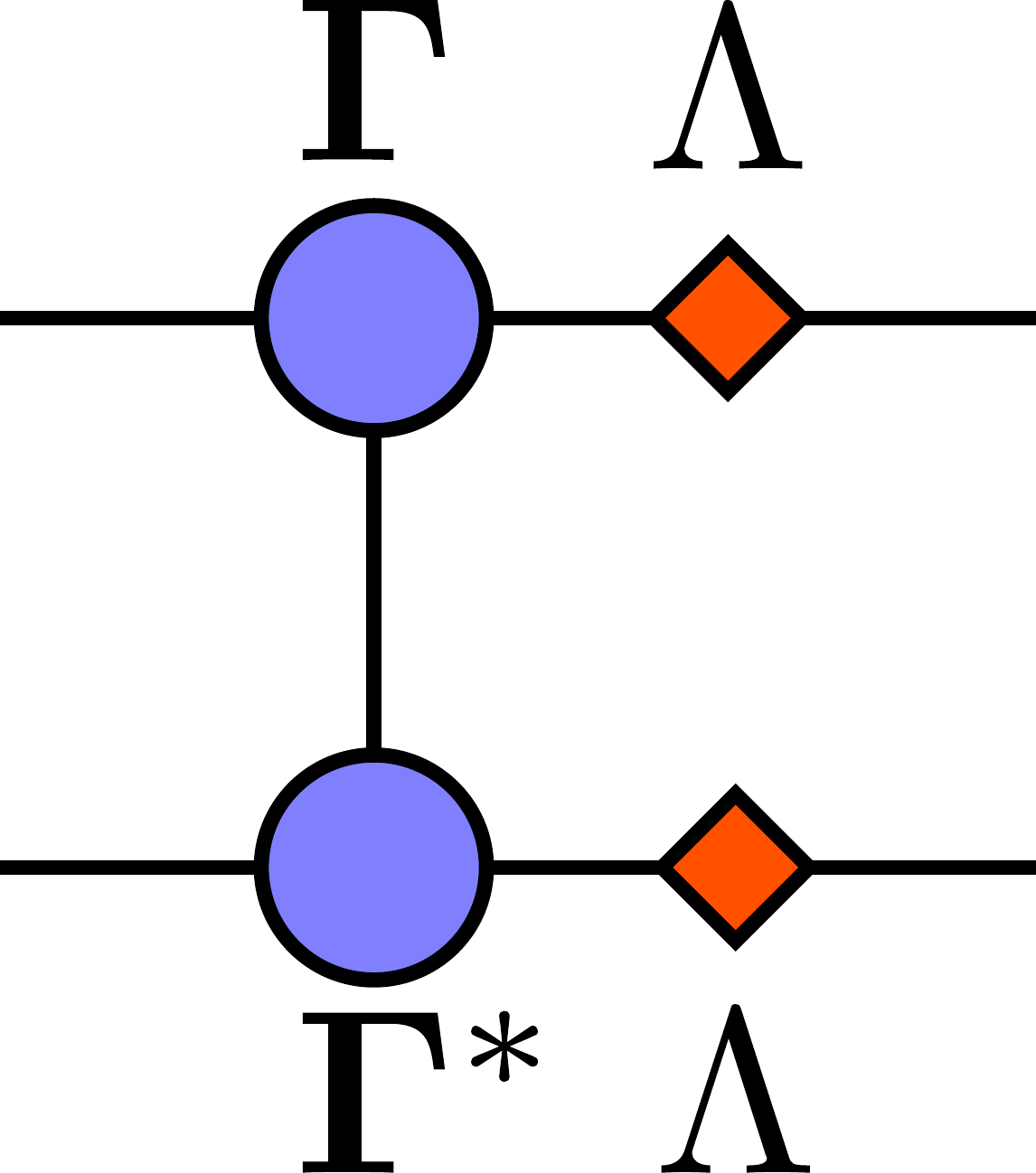} 
\end{equation}
to the eigenvalue $g=1$. We have numerically verified that $\bm 1$ is the only eigenvector of $\mathcal{T}$ with eigenvalue $|g| = 1$.

The $\omega_{12}$ quantum number is determined by the momentum per unit length of $\ket{s}$ in Eq.~(\ref{eq:mps}). It is thus given by the dominant eigenvalue $g_{T_2}$ of the generalized transfer matrix

\begin{equation}
	\label{eq:TMomega12}
	\begin{split}
		& \mathcal{T}_{T_2} \ \equiv \ 
		\ingr{1.772142}{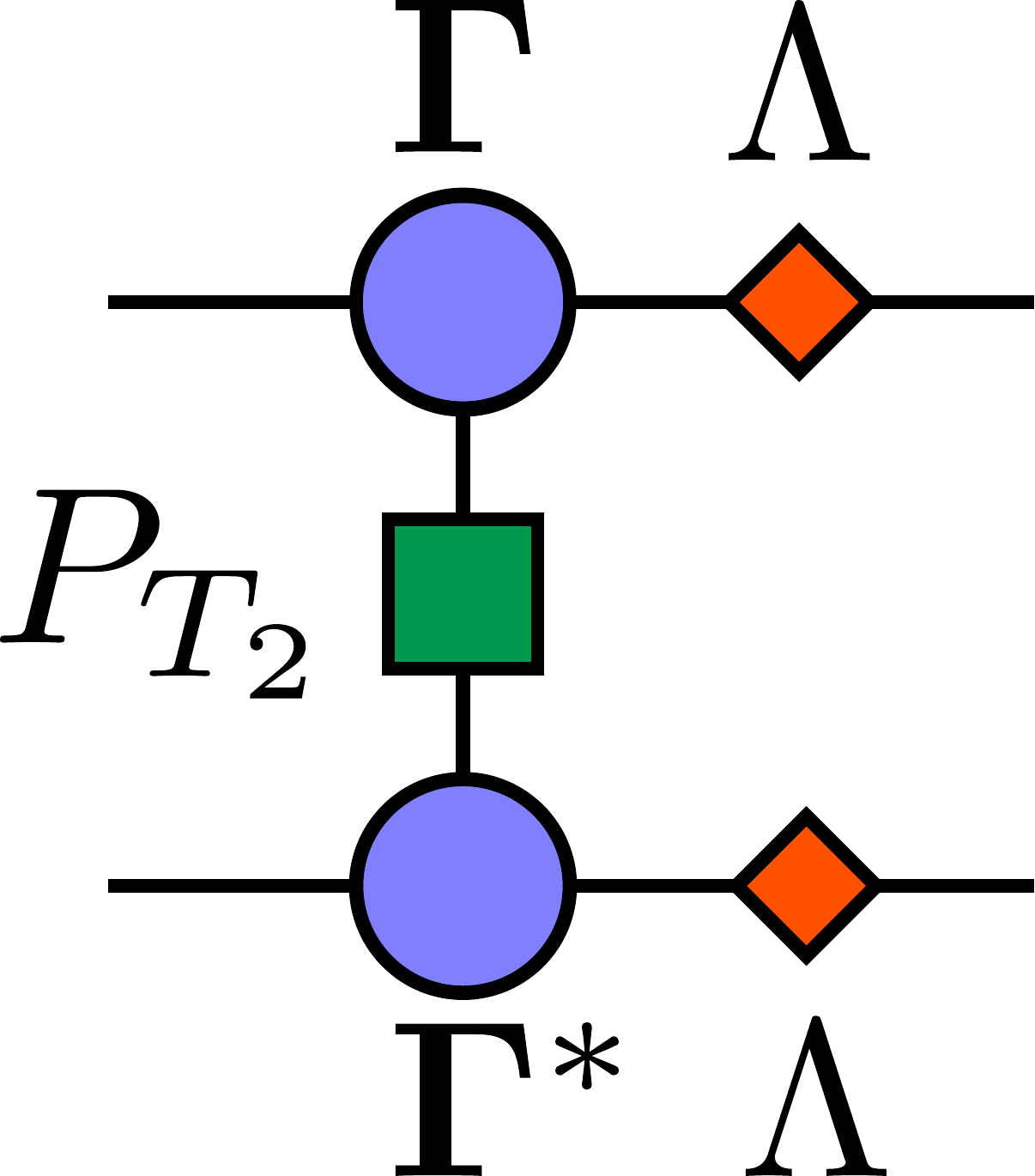} \ = \\ 
		& \ingr{1.772142}{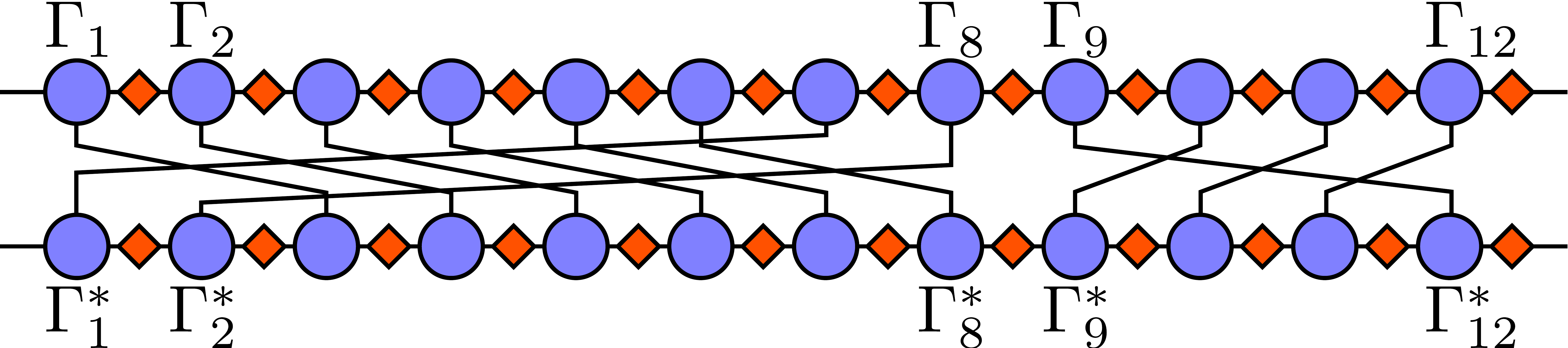} \ ,
	\end{split}
\end{equation}
where $P_{T_2}$ is a permutation of lattice sites within the MPS unit cell that corresponds to a translation in $\hat{y}$ direction by one lattice unit cell (see Fig.~\ref{fig:XC8-4}). We stress that the translational invariance in $\hat{y}$ direction is not built in the MPS ansatz for a ground state on an infinite cylinder. It is only (approximately) recovered as the energy optimization of tensors in Eq.~(\ref{eq:mps}) reaches convergence. We have observed numerically that the dominant eigenvalue $g_{T_2} = -1 $ for a converged ground state. From this we conclude that $w_{12} = -1$, in agreement with prediction of Sec.~\ref{sec:classify-csf}.

We now turn to the computation of $\omega_{\sigma^\ast}$. A mirror
symmetry $\sigma$ is chosen such that it does not exchange left and
right sides of an infinite cylinder and is compatible with the MPS
covering of the cylinder, see Fig.~\ref{fig:XC8-4}. Note that $\sigma$ is an
internal symmetry of the ground state $\ket s$ in Eq.~(\ref{eq:mps})
if the MPS unit cell is enlarged to include $N=48$ lattice
sites. Since the ground state $\ket s$ is
invariant under the combination $\sigma^\ast = \sigma T$, the tensor
$\bm \Gamma$ in Eq.~(\ref{eq:mps}) transforms in the following
way\cite{Pollmann2010}
\begin{equation}
	\label{eq:Gsigma}
	\ingr{1.618890}{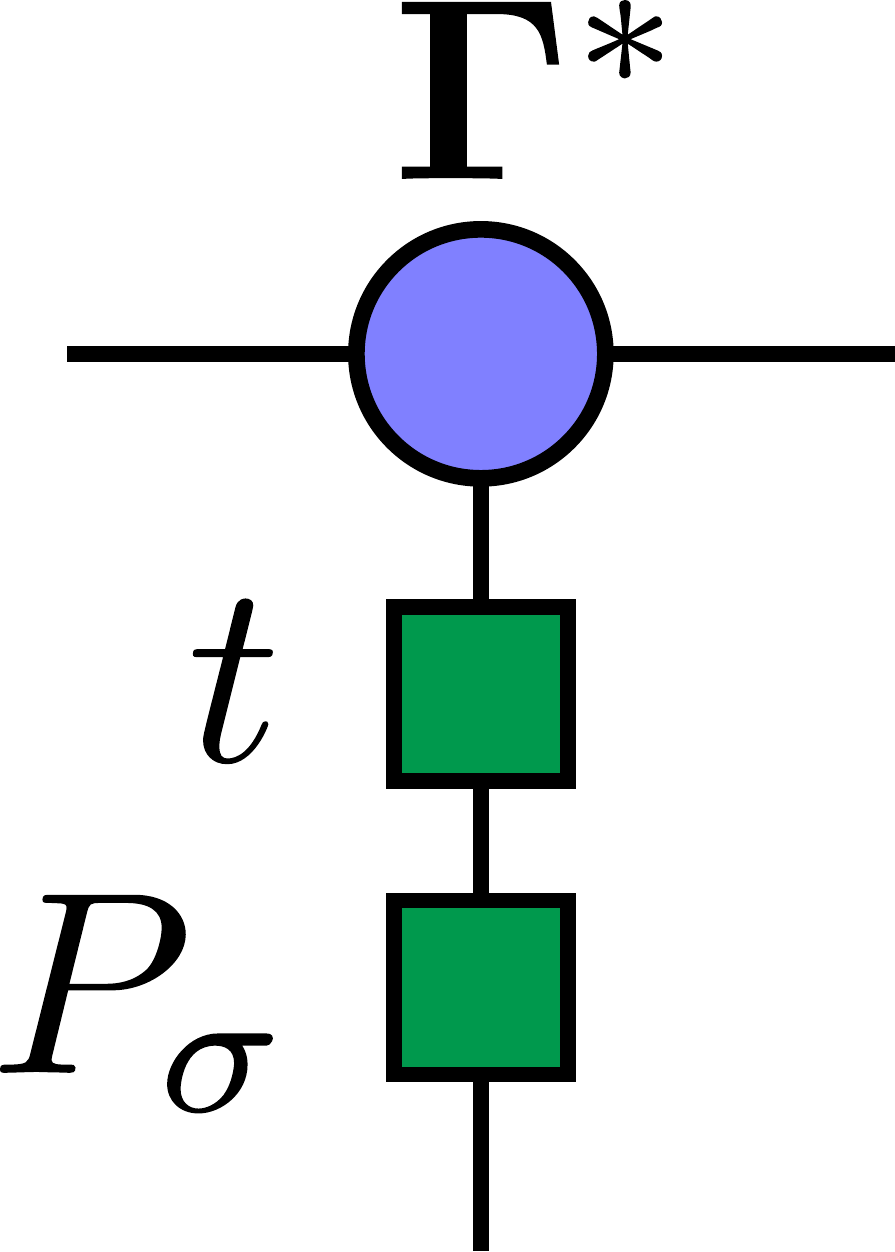} \ = \ e^{i\theta_{\sigma^\ast}} \ \cdot \
	\ingr{1.095048}{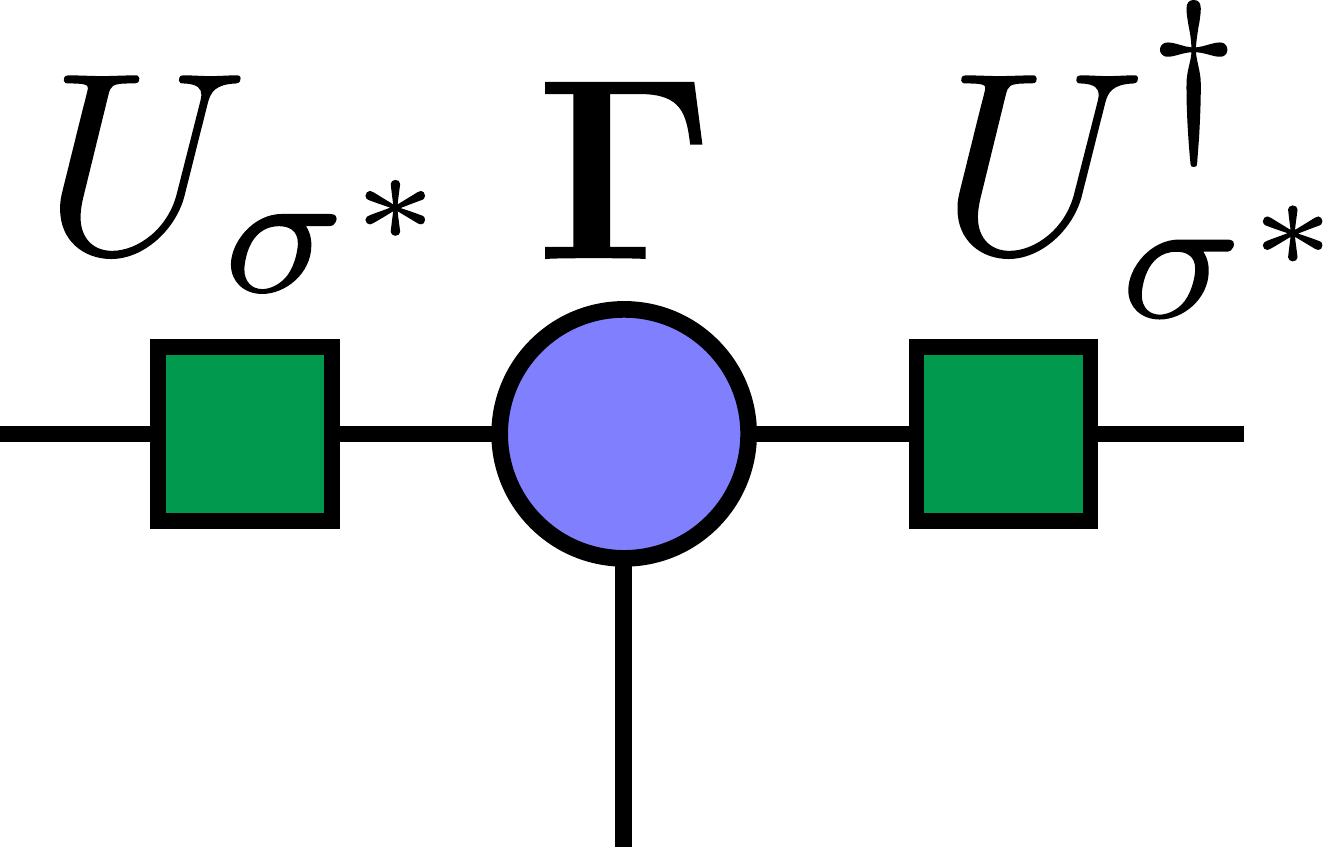} \ ,
\end{equation}
where $t \equiv \prod_j e^{i\pi S^y_j}$ together with conjugating the tensor $\bm \Gamma$ implements the time-reversal operation. $P_\sigma$ is a permutation of lattice sites within (enlarged) MPS unit cell that corresponds to a mirror reflection $\sigma$. $U_{\sigma^\ast}$ is a unitary matrix that commutes with $\Lambda$ in Eq.~(\ref{eq:mps}) and $\theta_{\sigma^\ast}$ is a non-universal phase. It follows from Eq.~(\ref{eq:Gsigma}) that the matrix $U_{\sigma^\ast}$ is the leading right eigenvector of the generalized transfer matrix
\begin{equation}
	\label{eq:TMsigma}
	\mathcal{T}_{\sigma^\ast} \ \equiv \ 
	\ingr{2.153091}{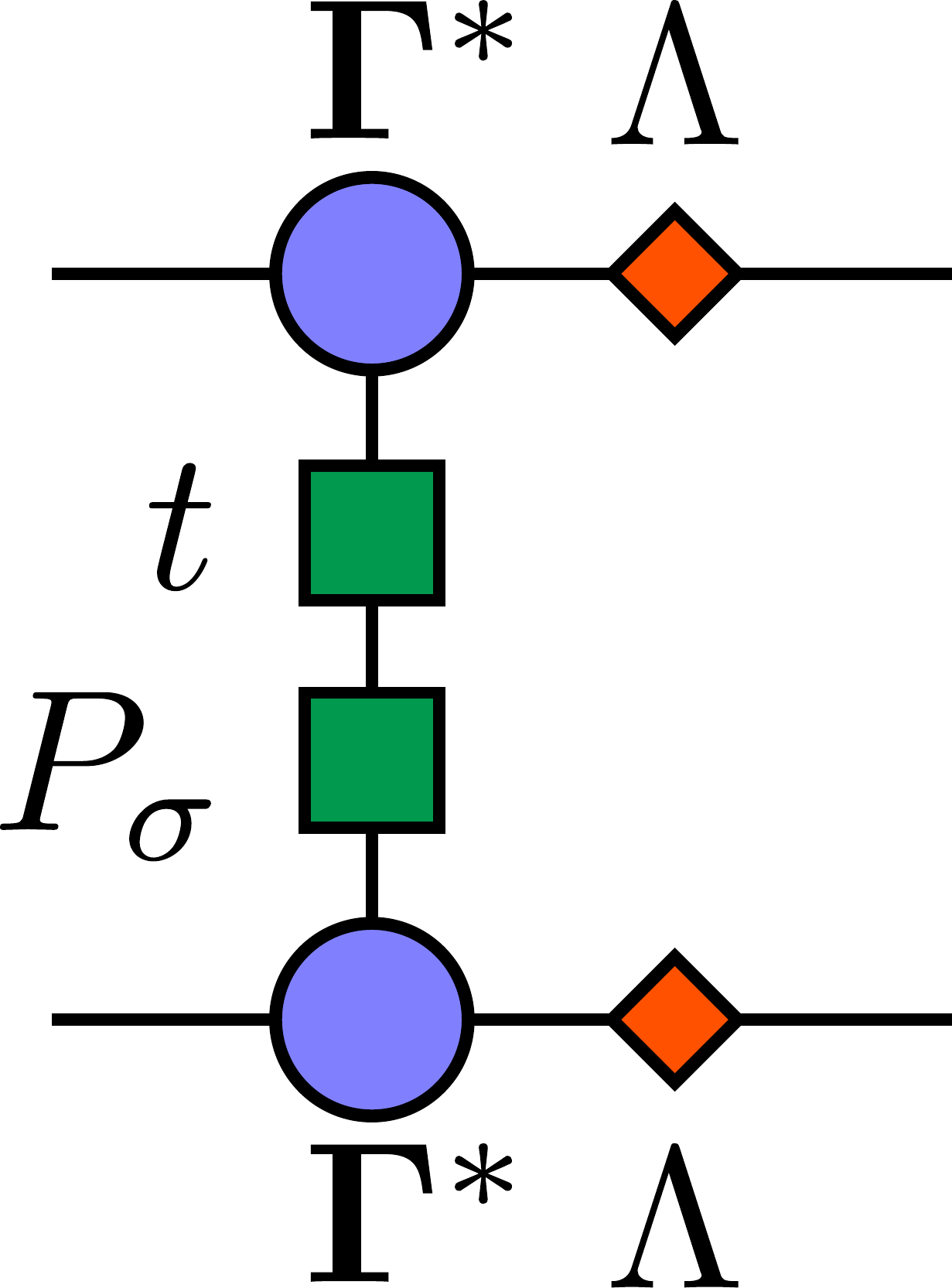} \ .
\end{equation}

By iterating Eq.~(\ref{eq:Gsigma}) twice, it can be shown that $U_{\sigma^\ast}^\dagger U_{\sigma^\ast}^{\mathsmaller T}$ is the right eigenvector of the transfer matrix $\mathcal{T}$ in Eq.~(\ref{eq:TM}) to the eigenvalue $g=1$. As noted before, identity $\bm 1$ is the only eigenvector of $\mathcal{T}$ to the eigenvalue $g=1$ which means that $U_{\sigma^\ast}^\dagger U_{\sigma^\ast}^{\mathsmaller T} = \alpha_{\sigma^\ast} \bm 1$. Consequently, $U_{\sigma^\ast} = \alpha_{\sigma^\ast} U_{\sigma^\ast}^{\mathsmaller T}$, where $\alpha_{\sigma^\ast} = \pm 1$ is a 1D SPT invariant that corresponds to the $\omega_{\sigma^\ast}$ quantum number. Note that Eq.~(\ref{eq:Gsigma}) defines $U_{\sigma^\ast}$ only up to a phase. It has no impact on the above reasoning, since any phase ambiguity is canceled in the expression $U_{\sigma^\ast}^\dagger U_{\sigma^\ast}^{\mathsmaller T}$, as emphasized in Sec.~\ref{sec:csf-anyon}.

We have computed $U_{\sigma^\ast}$ as the dominant right eigenvector of the transfer matrix $\mathcal{T}_{\sigma^\ast}$ in Eq.~(\ref{eq:TMsigma}) and verified that $U_{\sigma^\ast} = - U_{\sigma^\ast}^{\mathsmaller T}$ which shows that $\omega_{\sigma^\ast} = -1$, as expected.

The computation of $\omega_{\mu^\ast}$ is performed in a similar
manner. First, the ground state $\ket{s}$ on an infinite XC8 cylinder carrying
semion flux is found. We choose mirror symmetry $\mu$ such
that it does not exchange left and right sides of the cylinder, see Fig.~\ref{fig:XC8}. The
MPS unit cell consists of $N=12$ tensors and it is organized in a way
that $\mu$ is an internal symmetry of the ground state $\ket s$ in
Eq.~(\ref{eq:mps}). The quantum number $\omega_{\mu^\ast}$ is deduced
by studying the symmetry properties of the right leading eigenvector
$U_{\mu^\ast}$ of the generalized transfer matrix
\begin{equation}
	\label{eq:TMmu}
	\mathcal{T}_{\mu^\ast} \ \equiv \ 
	\ingr{2.153091}{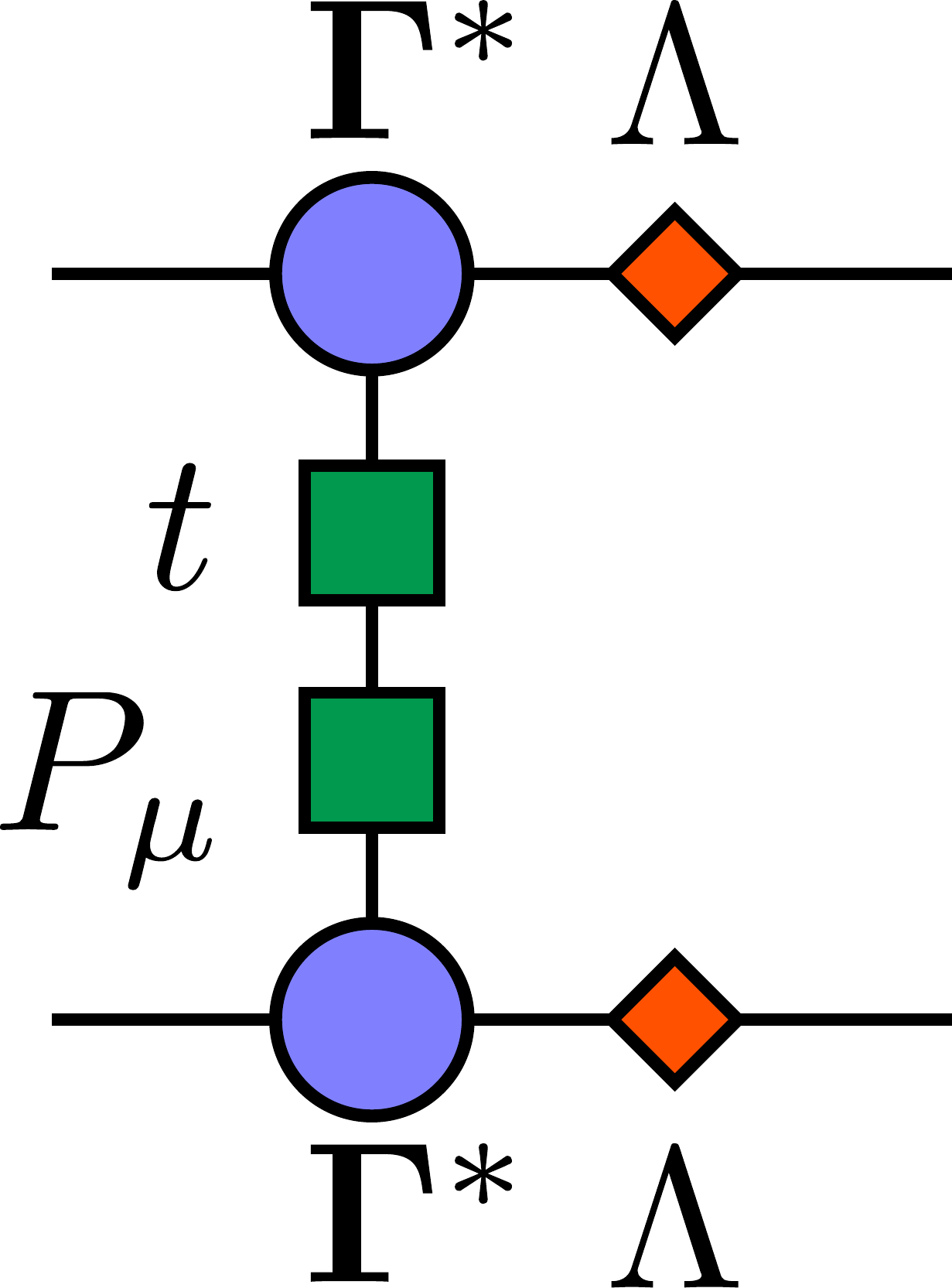} \ ,
\end{equation}
where $P_\mu$ is a permutation of sites that implements the mirror reflection $\mu$. We find that $U_{\mu^\ast}$ is antisymmetric, which shows that $w_{\mu^\ast} = -1$.

In order to conclude the detection of symmetry fractionalization of a chiral spin liquid under investigation, we extract $\omega_I$ quantum number from the ground state $\ket{s}$ on an XC8-4 infinite cylinder. The inversion $I$ maps an MPS unit cell to itself, provided that $\Gamma_j$ and $\Lambda_j$ in Eq.~(\ref{eq:MPSUnitCell}) are ``shifted'' by two positions along an MPS path, see Fig.~\ref{fig:XC8-4}. For a symmetric ground state $\ket{s}$, the tensor $\bm \Gamma$ in Eq.~(\ref{eq:mps}) transforms in the following way
\begin{equation}
	\label{eq:Ginver}
	\ingr{1.258646}{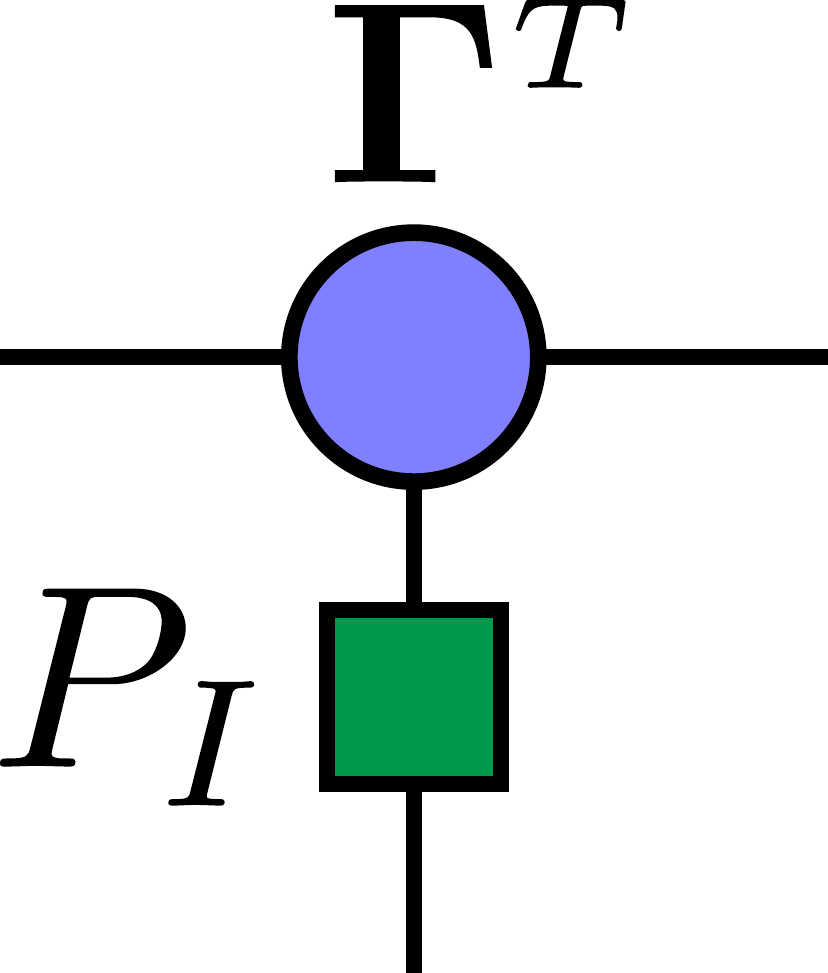} \ = \ e^{i\theta_I} \ \cdot \ 
	\ingr{1.095048}{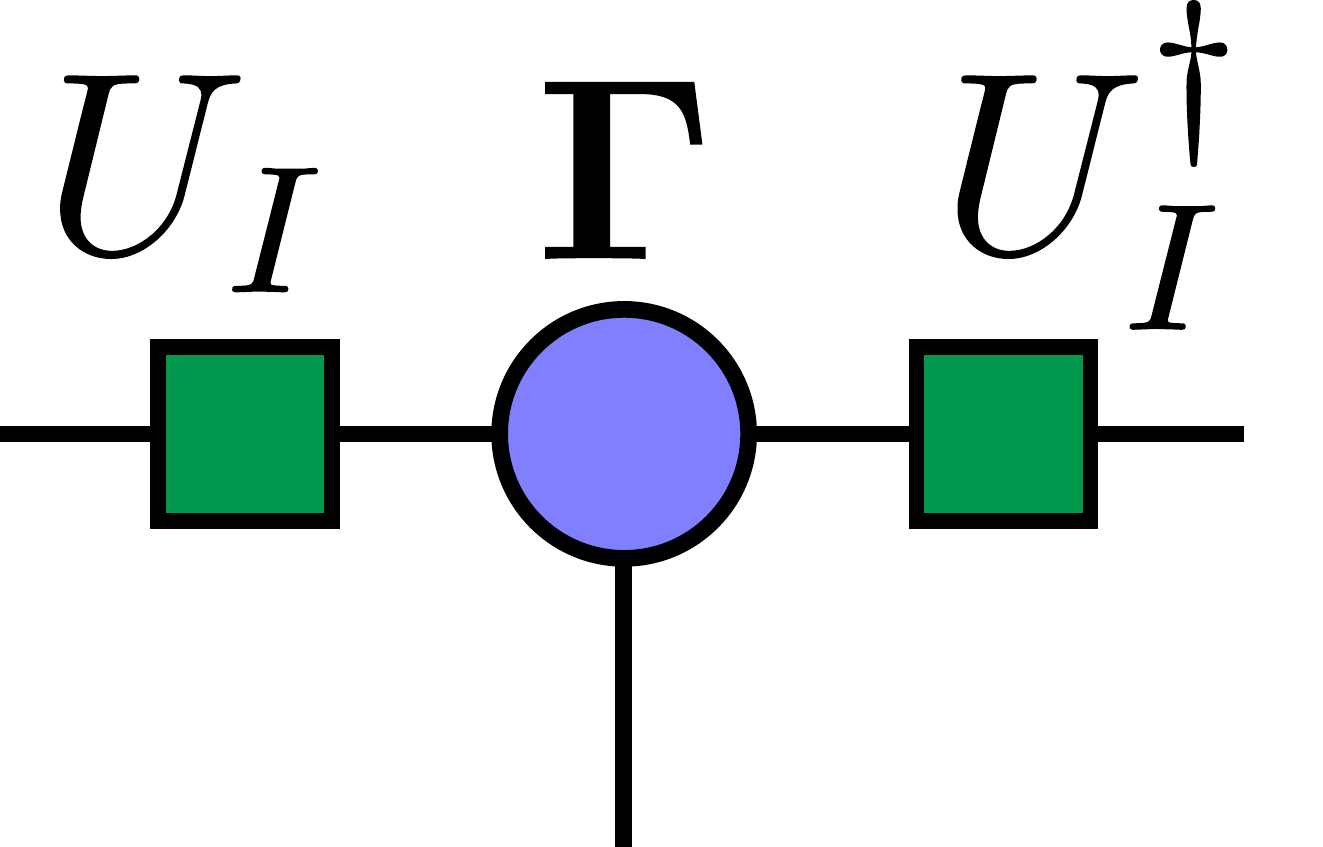} \ ,
\end{equation}
where $P_I$ is a permutation of lattice sites within a ``shifted'' MPS unit cell that corresponds to the inversion $I$. $U_I$ is a unitary matrix, $[U_I, \Lambda] = 0$, and $\theta_I$ is a phase. In close analogy to the case of mirror symmetry $\sigma$, it can be shown that $U_I$ is the leading right eigenvector of the generalized transfer matrix
\begin{equation}
	\label{eq:TMinver}
	\mathcal{T}_I \ \equiv \ \ingr{1.862662}{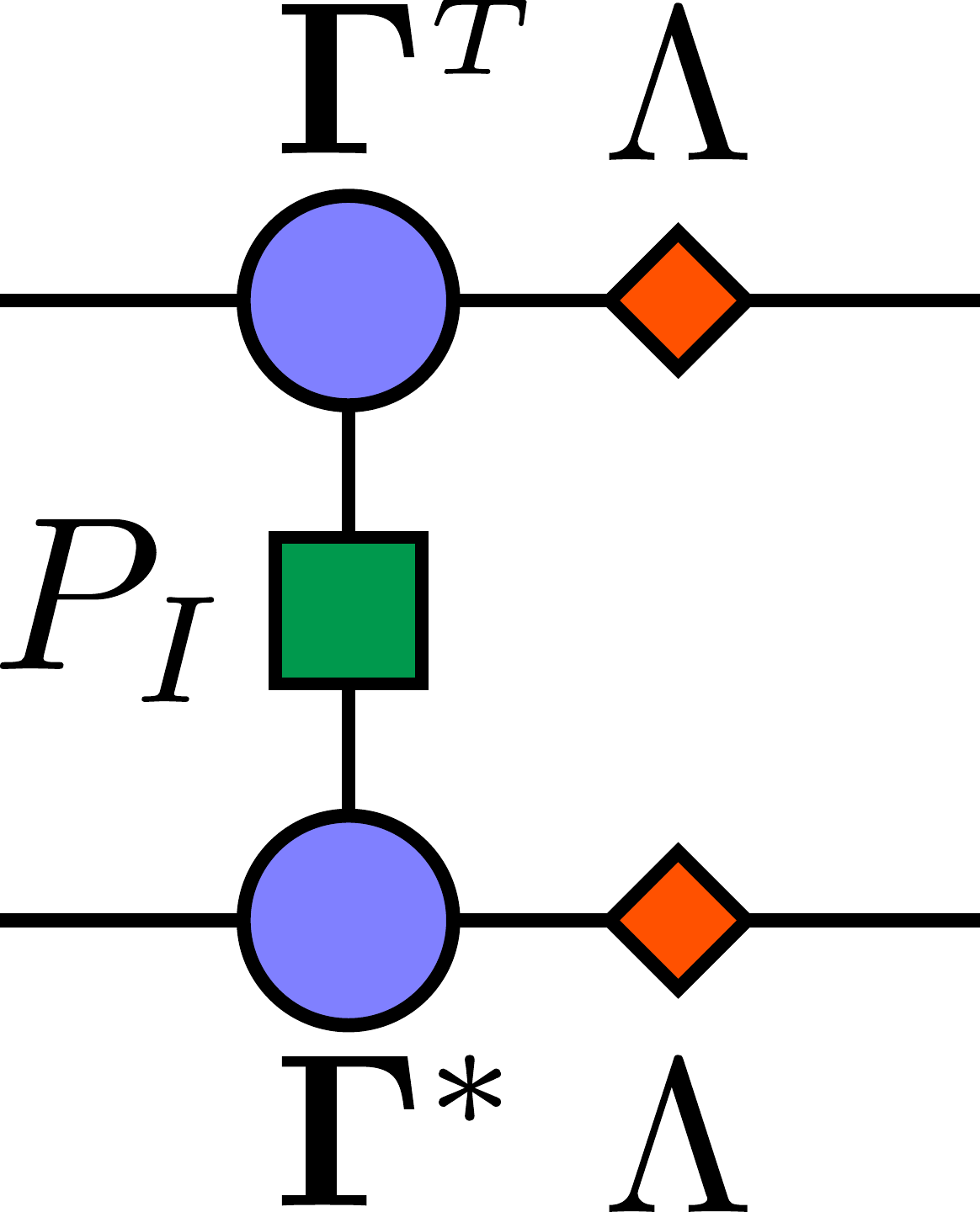}
\end{equation}
and that $U_I = \alpha_I U_I^{\mathsmaller T}$, where $\alpha_I = \pm 1$ is a 1D SPT invariant that corresponds to the $\omega_I$ quantum number. We have numerically computed $U_I$ and checked that it is anti-symmetric confirming that $\omega_I = -1$, as found in Sec.~\ref{sec:classify-csf}.

\section{Conclusion}
\label{sec:conclusion}

In this work, we classified and numerically detected the symmetry
fractionalization in chiral spin liquid states in a spin-$\frac12$
model on the kagome lattice. Using the recently proposed method of
flux-fusion anomaly test~\cite{HermeleFFAT}, we show that there are
only one anomaly-free way to fractionalize crystal symmetries in a
chiral spin liquid state, if we assume that there is a full SU(2)
spin-rotational symmetry and there is an odd number of spin-$\frac12$
per unit cell of the kagome lattice.

Our theoretical argument proves that all symmetric chiral spin liquids
found in previous numerical and theoretical studies belong to the
same SET phase. In particular, we provide numerical evidence that the
ground states obtained by DMRG belong to this SET phase. Moreover,
 we show in Appendix~\ref{sec:psg-af} that the chiral spin liquid state
obtained in the parton construction~\cite{MeiCSLX, HuCSL2015} also
belongs to the same phase.

In this work, we also provide general definitions for symmetry
fractionalization of space-time symmetries for Abelian anyons with
fractional statistics. In particular, the definition we provide for
unitary crystal symmetries applies consistently to generic Abelian
anyons, but differs from the previously adapted definition for the
special case of fermionic anyons by a minus sign. Our definition can
be used to study symmetry fractionalization in more complicated
Abelian topological orders, and we will leave this to future works.

\begin{acknowledgements}
We thank Xiao-Gang Wen, Jia-Wei Mei and Meng Cheng for enlightening discussions. L.C. acknowledges support by the John Templeton Foundation. This research was supported in part by Perimeter Institute for Theoretical Physics. Research at Perimeter Institute is supported by the Government of Canada through Industry Canada and by the Province of Ontario through the Ministry of Research and Innovation.

Note added: Upon completion of this paper, we became aware of an independent work~\cite{ZaletelCSLPSG}.
\end{acknowledgements}

\appendix

\section{PSG analysis in Abrikosov-fermion construction of chiral spin
  liquid.}
\label{sec:psg-af}

In this appendix we study the symmetry fractionalization in a chiral
spin liquid state on the kagome lattice obtained in fermionic parton
construction~\cite{WenCSL, MeiCSLX, HuCSL2015}, using the method of
projective symmetry group (PSG) analysis~\cite{wenpsg}. This approach
reproduces the symmetry fractionalization listed in
Table~\ref{tab:omega}, which is argued to be the only possibility in
Sec.~\ref{sec:classify-csf} of the main text.

It is well known that the chiral spin liquid, having the same
topological order as a $\nu=1/2$ FQH state~\cite{KalmeyerCSL1987},
can be obtained in a parton construction if the fermionic parton forms
a Chern insulator with the Chern number $n=2$~\cite{WenCSL}. On the
kagome lattice chiral spin liquids have been constructed using this
approach in Ref.~\onlinecite{MeiCSLX, HuCSL2015} and they have been
shown to have good variational energies in VMC
studies~\cite{HuCSL2015}.

\begin{figure}[htbp]
  \centering
  \subfigure[\label{fig:parton:0}]{\includegraphics{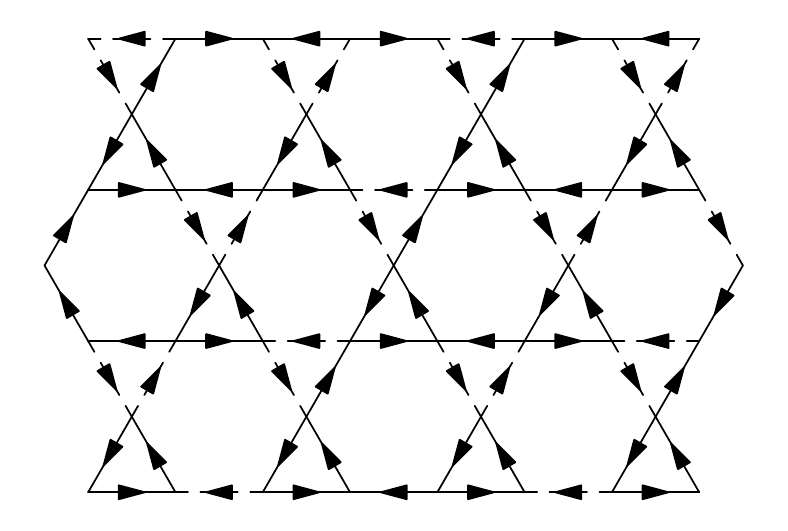}}
  \subfigure[\label{fig:parton:mu}$\mu^\ast$]{\includegraphics{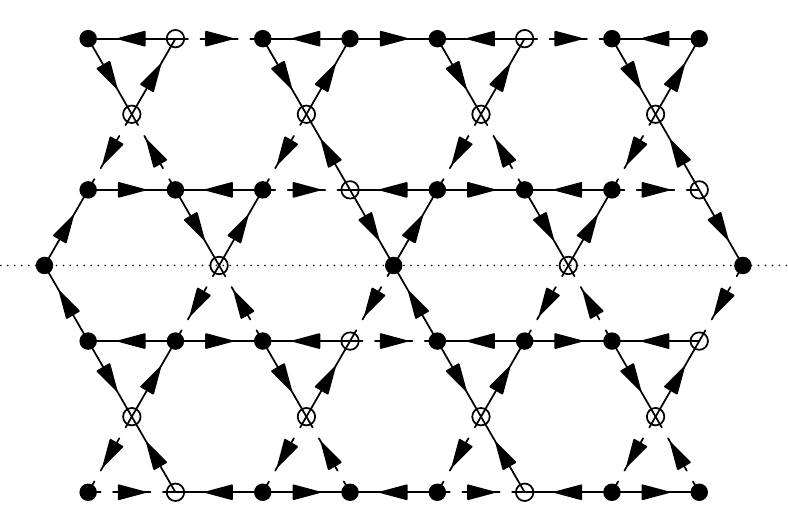}}
  \subfigure[\label{fig:parton:sigma}$\sigma^\ast$]{\includegraphics{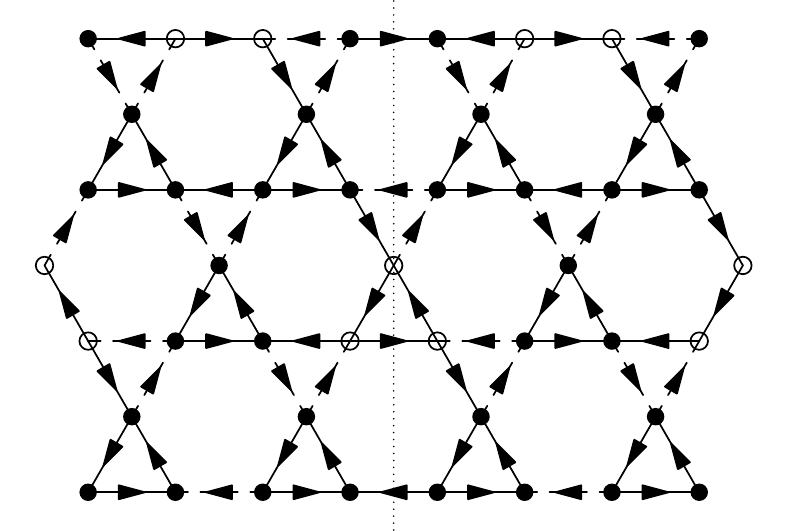}}
  \caption{Mean field ansatz and its PSG. Solid bonds and dashed bonds
    indicate hopping amplitudes $t_0$ and $-t_0$, respectively.
    The arrows on the bonds mark the direction along which the
    hopping is $\pm t_0$, while the hopping against the direction of
    the arrows is $\pm t_0^\ast$. (a) The nearest-neighbor hopping
    amplitudes in the mean field ansatz in
    Ref.~\onlinecite{HuCSL2015}. (b) The hopping amplitudes after the
    symmetry operation $\mu^\ast$. (c) The hopping amplitudes after
    the symmetry operation $\sigma^\ast$. In (b) and (c), the mirror
    axis is marked by a dotted line. The hopping amplitudes can be
    restored to the ones in (a) by a gauge transformation in
    Eq.~\eqref{eq:muT-Hu}, where $\eta_i=\pm1$ on sites marked by
    filled (open) circles, respectively.}
  \label{fig:parton}
\end{figure}

Here we choose the parton wave function constructed
by~\citet{HuCSL2015} as an example and study its symmetry
fractionalization using PSG analysis. In this construction the
spin-$\frac12$ degree of freedom on each site is represented by a
fermionic parton,
\begin{equation}
  \label{eq:spinon}
  \bm S_i=f_{i\alpha}^\dagger\bm\sigma_{\alpha\beta}f_{i\beta}.
\end{equation}
The wave function of a chiral spin liquid is then constructed by
Gutzwiller-projecting a mean field parton wave function,
\begin{equation}
  \label{eq:gp}
  |\Psi\rangle = P_G|\Psi_f\rangle,
\end{equation}
where $P_G$ denotes the Gutzwiller projection, which enforces the
constraint that there is exactly one parton per site, and
$|\Psi_f\rangle$ is the mean field parton wave function, which is the
ground state of the mean field Hamiltonian,
\begin{equation}
  \label{eq:Hmf}
  H_{\text{MF}} = \sum_{ij}\left(t_{ij}f_{i\alpha}^\dagger
    f_{j\alpha}+\text{h. c.}\right).
\end{equation}
The hopping amplitudes $t_{ij}$ are given in Fig.~1(a) of
Ref.~\onlinecite{HuCSL2015} and are shown in Fig.~\ref{fig:parton:0}
here (here we only show the nearest-neighbor hoppings because these
are enough to fix the PSG and the next-nearest-neighbor only improves
the variational energy without affecting the PSG). Note that because
the chiral spin liquid state breaks time-reversal symmetry, the
hopping amplitudes $t_{ij}$ are complex (an complex hopping flux is
required to obtain the nontrivial Chern number).

The parton representation given by Eq.~\eqref{eq:spinon} has an SU(2)
gauge structure~\cite{AffleckSU21988, LNWReview}, which is broken to
U(1) by the mean field Hamiltonian in Eq.~\eqref{eq:Hmf}. In other
words, the invariant gauge group (IGG)~\cite{wenpsg} of this parton
wave function is U(1).

In this construction the IGG does not match the topological order. As
a result the projective representation the fermionic parton carries
does not contain all information of the projective representation
carried by the semion. Particularly the former is described by
$H^2(G, \mathrm U(1))$ where the coefficient is given by the
$\mathrm{IGG}=\mathrm U(1)$, while the latter is described by
$H^2(G, \mathbb Z_2)$, where the coefficient is $\mathbb Z_2$ because
of the fusion rule $s\times s=1$. In comparison, in parton
constructions of a $\mathbb Z_2$ spin liquid~\cite{wenpsg} both the
IGG and the fusion rule of the fermionic anyon are $\mathbb Z_2$, so
there is a one-to-one correspondence between the projective
representations carried by the fermionic parton and the fermionic
anyon.

The quantum number fractionalization $X^2=\pm1$ of an antiunitary
symmetry operator $X$ can still be extracted from the PSG, because
$H^2(\mathbb Z_2^X, \mathrm U(1))=H^2(\mathbb Z_2^X, \mathbb
Z_2)=\mathbb Z_2$. Namely this symmetry fractionalization can be
extracted from how $X$ acts on the parton operator
$X^2f_{i\alpha}=\omega_Xf_{i\alpha}$, because for an antiunitary
operator $X$ the $\omega_X$ obtained this way is invariant under a
U(1) gauge transformation and distinguishes different PSGs. For the
mean field ansatz presented in Ref.~\onlinecite{HuCSL2015}, the
parton $f_{i\alpha}$ carries the following projective representation
of $\mu^\ast=\mu T$,
\begin{equation}
  \label{eq:muT-Hu}
  \mu^\ast: f_{i\alpha}\rightarrow
  \eta_{\mu^\ast}(i)\epsilon_{\alpha\beta}f_{\mu(i)\beta},
\end{equation}
where $\mu(i)$ denotes the site that is the mirror image of site $i$,
and $\eta_{\mu^\ast}(i)$ is a site-dependent phase factor. Up to a
U(1) gauge transformation (which does not affect the result of
$\omega_{\mu^\ast}$), the choice of $\eta_{\mu^\ast}(i)$ shown in
Fig.~\ref{fig:parton:mu} leaves the mean field Hamiltonian in
Eq.~\eqref{eq:Hmf} invariant. It is easy to check this choice
satisfies
$\eta_{\mu^\ast}(\mu(i))\eta_{\mu^\ast}^\ast(i)=+1$. Therefore from
Eq.~\eqref{eq:muT-Hu} we can derive that
$(\mu^\ast)^2f_{i\alpha}=-f_{i\alpha}$, which is consistent with the
conclusion of $\omega_{\mu^\ast}=-1$ obtained in the main text as listed
in Table~\ref{tab:omega}.

Similarly the projective representation of $\sigma^\ast=\sigma T$
carried by the parton $f_{i\alpha}$ has a similar form of
Eq.~\eqref{eq:muT-Hu}, with another site-dependent factor
$\eta_{\sigma^\ast}(i)$, as shown in Fig.~\ref{fig:parton:sigma}. This
factor also satisfies
$\eta_{\sigma^\ast}(\sigma(i))\eta_{\sigma^\ast}^\ast(i)=+1$, and this
implies $\omega_{\sigma^\ast}=-1$, which is also consistent with the
conclusion in Table~\ref{tab:omega} of the main text.

Furthermore, the commutation relation fractionalization
$T_1T_2=\pm T_2T_1$ can also be determined from the parton PSG. It is
well known that the parton carries a nontrivial commutation relation
fractionalization $T_1T_2=-T_2T_1$ as the parent gapless U(1) spin
liquid state is known as a $\pi$-flux state~\cite{Hermele2008},
indicating that the hopping amplitudes of the fermionic parton has a
$\pi$-flux in each unit cell (this can be easily checked from
Fig.~\ref{fig:parton}). In the main text we argued that this is the only possibility
in the chiral spin liquid state.

However, the symmetry fractionalization $I^2=\omega_I=\pm1$ (see the
definition in Table~\ref{tab:omega} of the main text) cannot be directly
derived from the parton PSG, because the parton does not have a
nontrivial projective representation of inversion as $H^2(\mathbb Z_2,
\mathrm U(1))$ is trivial. In other words, as the IGG is U(1), the
parton's projective representation of inversion has a U(1) phase
ambiguity $e^{i\theta}$ and correspondingly the action of $I^2$ has a
phase ambiguity of $e^{2i\theta}$. Thus the two projective
representations $I^2=\pm1$ are smoothly connected and cannot be
distinguished from each other by studying parton PSG.

This symmetry fractionalization $\omega_I$, however, can be obtained
from this parton construction by studying the inversion eigenvalue of
the ground state wave function. In Sec.~\ref{sec:csf-anyon} of the
main text we showed that $\omega_I$ can be defined as the inversion
parity eigenvalue of the two-semion wave function. Because the model
has one spin-$\frac12$ on each site, the chiral spin liquid state has
a nontrivial background anyon charge of one semion per
site~\cite{Zaletel2015, ChengLSM}, and consequently $\omega_I$ can
also be detected from the inversion eigenvalue of the ground state
wave function on a system with $(4n+2)$ sites and open boundaries. In
this parton construction, such a wave function is obtained by
projecting a mean field wave function as in Eq.~\eqref{eq:gp}, where
the mean field wave function, as the ground state of the mean field
Hamiltonian, can be obtained by filling all eigenmodes of the
Hamiltonian that lies below the Fermi energy,
\begin{equation}
  \label{eq:Psif-mf}
  |\Psi_f\rangle=\prod_{\lambda,E_\lambda<0} 
  f_{\lambda\uparrow}^\dagger f_{\lambda\downarrow}^\dagger|0\rangle,
\end{equation}
where $\lambda$ labels all eigenmodes of the free mean field
Hamiltonian in Eq.~\eqref{eq:Hmf} and the product runs over all
$\lambda$ with a negative energy $E_\lambda<0$. Note that the
Hamiltonian in Eq.~\eqref{eq:Hmf} does not act on the spin quantum
number, so all modes are doubly degenerate with $\alpha=\uparrow$ and
$\downarrow$.

Next, we consider the action of inversion symmetry on the mean field
Hamiltonian. We assume that the fermion takes the following projective
representation of $I$ denoted as $\tilde I$,
\begin{equation}
  \label{eq:Itilde}
  \tilde I: f_{i\alpha}\rightarrow \eta_I(i)f_{I(i),\alpha},
\end{equation}
where the phases $\eta_I(i)$ satisfies
\begin{equation}
  \label{eq:etaI}
  \eta_I(i)\eta_I(I(i))=e^{i\Theta}.
\end{equation}
Here $e^{i\Theta}$ is a site-independent phase factor, and the
projective representation satisfies $\tilde I^2=e^{i\Theta}$. As we
discussed before, one can choose different forms of the projective
representation $\tilde I$ satisfying $\tilde I^2=e^{i\Theta}$ with
arbitrary phase $e^{i\Theta}$, but this phase ambiguity does not
affect the result of symmetry fractionalization $I^2=\pm1$ we will
derive below.

In the mean field theory the symmetry $\tilde I$ can be viewed as a
global symmetry under which the parton $f_{i\alpha}$ and wave
functions all carry linear representations. Particularly
$\tilde I H\tilde I^{-1}=H$ so we assume the eigenmodes are also
eigenstates of $\tilde I$. Furthermore, because the inversion symmetry
also commutes with spin rotations, the two degenerate modes
$f_{\lambda\uparrow}$ and $f_{\lambda\downarrow}$ have the same
inversion eigenvalue
$\tilde If_{\lambda\alpha}=\Lambda_\lambda f_{\lambda\alpha}$. If
$\tilde I^2=e^{\Theta}$, $\Lambda_\lambda$ can only take one of the
two values $\Lambda_\lambda=\pm e^{i\Theta/2}$. Therefore the
inversion eigenvalue of the mean field wave function in
Eq.~\eqref{eq:Psif-mf} is determined by multiplying all inversion
eigenvalues of occupied modes:
\begin{equation}
  \label{eq:IPsif}
  \tilde I|\Psi_f\rangle
  =\prod_{\lambda,E_\lambda<0}(\Lambda_\lambda)^2|\Psi_f\rangle
  =(e^{i\Theta})^{2n+1}|\Psi_f\rangle.
\end{equation}
In the last step we used the fact that there are in total $(4n+2)$
occupied modes.

Next we study how $\tilde I$ acts on the Gutzwiller projection $P_G$. We saw
that the $\tilde I$ eigenvalue of the mean field wave function depends
on $e^{i\Theta}$, which in turn depends on the gauge choice in
Eq.~\eqref{eq:Itilde}. This U(1) gauge dependence is canceled by the
action of $\tilde I$ on $P_G$, as the physical spin wave function
obtained after the Gutzwiller projection should be gauge invariant. By
definition $P_G$ maps the Fock states of parton operator to spin
states,
\begin{equation}
  \label{eq:PG}
  P_G=\prod_i\sum_\alpha|\alpha\rangle_i\langle0|f_{i\alpha}.
\end{equation}
Therefore applying the transformation in Eq.~\eqref{eq:Itilde} we
conclude that $P_G$ transforms as the following under $\tilde I$,
\begin{equation}
  \label{eq:IPGI}
  \tilde IP_G\tilde I^{-1}=(-e^{i\Theta})^{2n+1}P_G,
\end{equation}
where the minus sign comes from exchanging the two fermionic parton
operators at $i$ and $I(i)$ after the inversion.

Combining the results in Eq.~\eqref{eq:IPsif} and \eqref{eq:IPGI}, we
obtain that that the eigenvalue of the physical spin wave function is
$(-1)^{2n+1}=-1$. Therefore using the definition we describe in
Sec.~\ref{sec:csf-anyon} of the main text we conclude that the semions
carry a nontrivial symmetry fractionalization $I^2=-1$.

\bibliography{csl}

\end{document}